\documentclass[twocolumn]{aastex62}
\usepackage{verbatim}
\usepackage{amstext}
\usepackage{amssymb}
\usepackage{graphicx}

\makeatletter

\providecommand{\tabularnewline}{\\}

\submitjournal{ApJ}

\shorttitle{First Galaxies in Peaks and Voids}
\shortauthors{Ahn \& Smith}

\begin{document}

\title{Formation of First Galaxies inside Density Peaks and Voids under
the Influence of Dark Matter-Baryon Streaming Velocity, I: Initial
Condition and Simulation Scheme}

\correspondingauthor{Kyungjin Ahn}
\email{kjahn@chosun.ac.kr}

\author{Kyungjin Ahn}
\affiliation{Department of Earth Sciences, Chosun University, Gwangju 61452, Korea}

\author{Britton D. Smith}
\affiliation{San Diego Supercomputer Center, University of California at San Diego,
San Diego 92093, USA}

\begin{abstract}
We present a systematic study of the cosmic variance that existed
in the formation of first stars and galaxies. We focus on the cosmic
variance induced by the large-scale density and velocity environment
engraved at the epoch of recombination. The density environment is
predominantly determined by the dark-matter overdensity, and the velocity
environment by the dark matter-baryon streaming velocity. Toward this
end, we introduce a new cosmological initial condition generator BCCOMICS,
which solves the quasi-linear evolution of small-scale perturbations
under the large-scale density and streaming-velocity environment and
generates the initial condition for dark matter and baryons, as either
particles or grid data at a specific redshift. We also describe a
scheme to simulate the formation of first galaxies inside density
peaks and voids, where a local environment is treated as a separate
universe. The resulting cosmic variance in the minihalo number density 
{\bf and} the amount of cooling mass are presented as
an application. Density peaks become a site for enhanced formation
of first galaxies, which compete with the negative effect from the
dark matter-baryon streaming velocity on structure formation.
\end{abstract}


\section{Introduction\label{sec:Introduction}}

First stars and first galaxies form out of the primordial environment
which is devoid of any metal, and they are categorized as Population
III (Pop III) objects. For such objects to form, at least high density,
efficient gas-phase cooling and shielding from the coolant-dissociating
radiation field are necessary. Such conditions can be first met inside
minihalos which are the first nonlinear objects in the universe. Pioneering
numerical simulations used to find that first stars are very massive
($M_{*}\gtrsim100-1000\,M_{\odot}$) and form in isolation inside
minihalos (\citealt{2002Sci...295...93A,2002ApJ...564...23B,Yoshida2006}),
but later, higher-resolution simulations found that multiple formation
of intermediate-mass ($M_{*}\lesssim30\,M_{\odot}$) stars were likely
as well (\citealt{Turk2009,Stacy2010,Greif2011a}). Semi-analytical
studies also find that intermediate- or even small-mass Pop III stars
may have formed out of the primordial environment (\citealt{Omukai2001,Nagakura2005}).
Different physical conditions inside minihalos likely lead to a wide
spectrum of the mass of Pop III stars (\citealt{Hirano2014,Hirano2015}).
Direct observation of truly metal-free stars is yet to be made in
future surveys. Current observations of ultra-iron-poor stars in the
Milky Way (e.g. \citealt{Caffau2011,Howes2015,Jacobson2015}), due
to the existence of other heavy elements, cannot be taken as a convincing
evidence of small-mass Pop III stars.

Because first stars and first galaxies form first inside minihalos,
varying physical conditions of minihalos (forming through the hydrogen-molecule
cooling) would lead to a variation in the outcome of the first star
formation. In addition to the minihalo-to-minihalo variance (\citealt{Hirano2014,Hirano2015}),
a large-scale variance in the streaming flow of baryons against dark
matter particles was found to impact the formation and evolution
of minihalos \citep[TH hereafter]{Tseliakhovich2010}. This effect
can be parameterized by the large-scale streaming velocity ${\bf V}_{{\rm bc}}\equiv{\bf V}_{{\rm b}}-{\bf V}_{{\rm {\rm c}}}$,
where ${\bf V}_{{\rm b}}$ and ${\bf V}_{{\rm c}}$ are the bulk velocities
of baryons and cold dark matter, respectively. In terms of the comoving
wavenumber, $k\simeq[10-1000]\,{\rm Mpc}^{-1}$ is the range strongly
affected, and results in the overall suppression of the matter density
fluctuation in this scale (TH). The coherence length of ${\bf V}_{{\rm bc}}$,
below which ${\bf V}_{{\rm bc}}$ hardly varies, is set by the baryon
acoustic oscillation (BAO) process during the epoch of recombination
and is in the order of a few comoving Mpc.

Subsequent studies have focused on both cosmological and astrophysical
implications of the impact of ${\bf V}_{{\rm bc}}$, and indeed have
found quantitative changes and new physical phenomena. Both the amplitude
and the peak location of the BAO feature, when observed through galaxy
surveys, will be affected mostly through the impact of ${\bf V}_{{\rm bc}}$
on the galaxy formation (\citealt{Dalal2010,Yoo2011,Blazek2016,Lewandowski2015,Slepian2015,Schmidt2016a}).
A new type of modulation on the 21-cm intensity mapping due to ${\bf V}_{{\rm bc}}$
and the corresponding boost of the signal may allow easier high-redshift
cosmology (\citealt{McQuinn2012}). The minimum mass of halos that
can host first stars will be increased from that without ${\bf V}_{{\rm bc}}$,
or first star formation will be delayed, due to the advection of gas
across the dark matter potential well, as found by numerical simulations
(\citealt{Greif2011,Maio2011,Stacy2011,O'Leary2012}). The mismatch
of CDM and baryon overdensities may induce baryon-dominated objects
such as globular clusters (\citealt{Naoz2014}) or yield the cosmic
variance in the gas content of halos (\citealt{Fialkov2012}). The
increase of the effective Jeans mass may induce the formation of large-mass,
direct-collapse black holes (\citealt{Hirano2017,Schauer2017,Regan2018}).

Unfortunately, numerical simulations mentioned above (\citealt{Greif2011,Maio2011,Stacy2011,O'Leary2012,Hirano2017,Schauer2017}),
except \citet{O'Leary2012}, are likely to underestimate the negative
effect of ${\bf V}_{{\rm bc}}$ on star formation and other similar
effects. This is because these work generated the initial condition
based on a typical Boltzmann code (such as CAMB) and imposed a sudden
${\bf V}_{{\rm bc}}$ at some initial time $t_{i}$. This procedure
would then completely underestimate the negative effect that had existed
before $t_{i}$. Instead, one should calculate the cumulative effect
of ${\bf V}_{{\rm bc}}$ from the recombination epoch to $t_{i}$
and then generate the initial condition, which would then contain
the negative effect that had existed before $t_{i}$. \citet{O'Leary2012}
followed this track by first calculating the evolution of perturbations
(given in terms of equations by TH) under the influence of ${\bf V}_{{\rm bc}}$
and then generating initial conditions, which were used for numerical
simulations of their nonlinear growth. As a result, they provided
a new initial condition generator CICsASS (Cosmological Initial Conditions
for AMR and SPH Simulations) that can be used to simulate the nonlinear
evolution of small-scale structures under a given ${\bf V}_{{\rm bc}}$
environment. 

In the meantime, improvements on the original formulation of TH was
made by considering long-range modes that had been neglected in TH
but found to be of higher significance than ${\bf V}_{{\rm bc}}$.
It was found that the large-scale overdensity, or equivalently the
velocity divergence, would impact the small-scale density fluctuation
more efficiently than ${\bf V}_{{\rm bc}}$ (\citealt[Ahn16
  hereafter]{Ahn2016}; \citealt{Blazek2016,Schmidt2016a}).
Ahn16 adopted the original peak-background split scheme of TH but
with additional large-scale mode contributions, approximating long-range
(small wavenumber) modes as a uniform local patch with given overdensities
($\Delta_{{\rm c}}$ and $\Delta_{{\rm b}}$), velocity divergences
($\Theta_{{\rm c}}$ and $\Theta_{{\rm b}}$), ${\bf V}_{{\rm bc}}$, and the temperature
fluctuation $\Delta_{T}$.
Ahn16 found that small-scale perturbations would evolve in a biased
way: the higher $\Delta_{\rm c}$ was, the faster $\delta$ would grow.
\citet{Schmidt2016a} showed that the galaxy bias and subsequently
their clustering are more strongly affected by the large-scale differential
overdensity ($\Delta{}_{{\rm bc}}\equiv\Delta_{{\rm c}}-\Delta_{{\rm b}}$,
the difference between the CDM overdensity $\Delta_{{\rm c}}$ and
the baryon overdensity $\Delta_{{\rm b}}$) and the differential velocity
divergence ($\Theta{}_{{\rm bc}}\equiv\Theta_{{\rm c}}-\Theta_{{\rm b}}$,
the difference between the CDM velocity divergence $\Theta_{{\rm c}}$
and the baryon velocity divergence $\Theta_{{\rm b}}$) than ${\bf V}_{{\rm bc}}$.
\citet{Blazek2016} invoked a similar perturbative approach as in
\citet{Schmidt2016a} including all 1-loop contributions, and found
that the BAO features (both the amplitude and the peak location) should
be affected. 


In order to study the growth of small-scale structures under the given
large-scale environment, characterized mainly of $\Delta$, $\Theta$
and ${\bf V}_{{\rm bc}}$, a most accurate way would be to (1) generate
an initial condition at redshift high enough to make perturbative
calculations valid and (2) then numerically simulate their evolution
to the nonlinear regime. For (1), even CICsASS falls short of the
requirement: initial conditions are adequately generated only for
mean-density ($\Delta=0$) regions in CICsASS because the original
formulation of TH is used. Ahn16 showed clearly that $\delta$ would
grow faster under larger $\Delta$ environment, and thus the full
set of equations treating large-scale impact, given by Ahn16, should
be integrated in order to generate an initial condition. This requirement
gets even stronger if one is interested in a very high-density
or a low-density environment,
because then the error from just using equations of TH instead of
Ahn16 would get larger. Here we introduce our initial condition generator
BCCOMICS\footnote{The code is available at https://www.github.com/KJ-Ahn/BCCOMICS.} 
(Baryon-CDM COsMological Initial Condition generator for
Small scales) that fulfills this requirement. For (2), because we
want to cover any overdensity environment, the usual set-up with zero
overdensity and a periodic boundary condition would not work. This
instead is allowed by using the well-known trick of treating the environment
as a separate universe with local cosmological parameters, which are
different from those of the global universe (e.g. \citealt{Goldberg2004}).

This paper is organized as follows. We describe the framework for
BCCOMICS in Section \ref{sec:BCCOMICS}, including the spatial variance
of large-scale environment, how this enters the evolution equation
of small-scale perturbations, and the schematics for generating initial
conditions from the transfer functions that become anisotropic. In
Section \ref{sec:simulating_peaks_voids} we describe a strategy to
identify an overdense (underdense) patch as a separate universe by
defining local parameters. In Section \ref{sec:Application}, as an
application of the schemes in Sections \ref{sec:BCCOMICS} and \ref{sec:simulating_peaks_voids},
we describe a suite of numerical simulations of high-redshift, small-scale
structure formation inside varying large-scale overdensity and ${\bf V}_{{\rm bc}}$
environments. In Section \ref{sec:Summary-and-Discussion}, we conclude
the paper with a summary. Unless noted otherwise, length scales are
expressed in comoving length units.

\section{Initial Condition: BCCOMICS\label{sec:BCCOMICS}}

In order to generate initial conditions for this study, we need an
initial condition generator that implements environmental effects from
both streaming-velocity and density. This requires first solving the
perturbation equations for the evolution of dark-matter and baryon
components at the least. In the early universe, perturbative description
of their evolution is justified when the scales of interest have not
entered the nonlinear regime yet.

The perturbation theory for the evolution of small-scale structures
under the influence of only the large-scale streaming velocity environment
was formulated by TH, and there exists an initial condition generator
that implements this theory (CICsASS, \citealt{O'Leary2012}). In terms of
overdensity ($\delta\equiv(\rho-\bar{\rho})/\bar{\rho}$, defined
with the local density $\rho$ and the average density $\bar{\rho}$),
peculiar velocity (${\bf v}$, with its divergence $\theta\equiv(1/a)\nabla\cdot{\bf v}$
where $a$ is the scale factor and $\nabla$ is the gradient in the
comoving frame) and temperature fluctuation ($\delta_{T}\equiv(T-\bar{T})/\bar{T}$,
with the local baryon temperature $T$ and the mean baryon temperature
$\bar{T}$), they solve the following linearized perturbation equation
in Fourier space:
\begin{eqnarray}
\frac{\partial\delta_{{\rm c}}}{\partial t} & = & -\theta_{{\rm c}},\nonumber \\
\frac{\partial\theta_{{\rm c}}}{\partial t} & = & -\frac{3}{2}H^{2}\Omega_{m}\left(f_{{\rm c}}\delta_{{\rm c}}+f_{{\rm b}}\delta_{{\rm b}}\right)-2H\theta_{{\rm c}},\nonumber \\
\frac{\partial\delta_{{\rm b}}}{\partial t} & = & -ia^{-1}{\bf V}_{{\rm bc}}\cdot{\bf k}\delta_{{\rm b}}-\theta_{{\rm b}},\nonumber \\
\frac{\partial\theta_{{\rm b}}}{\partial t} & = & -ia^{-1}{\bf V}_{{\rm bc}}\cdot{\bf k}\theta_{{\rm b}}-\frac{3}{2}H^{2}\Omega_{m}\left(f_{{\rm c}}\delta_{{\rm c}}+f_{{\rm b}}\delta_{{\rm b}}\right)-2H\theta_{{\rm b}}\nonumber \\
&&+a^{-2}\frac{k_{B}\bar{T}}{\mu m_{{\rm H}}}k^{2}\left\{ \delta_{T}+\delta_{{\rm b}}\right\} ,\nonumber \\
\frac{\partial\delta_{T}}{\partial t} & = & \frac{2}{3}\frac{\partial\delta_{{\rm b}}}{\partial t}-\frac{x_{e}(t)}{t_{\gamma}}a^{-4}\frac{\bar{T}_{\gamma}}{\bar{T}}\delta_{T},\label{eq:TH}
\end{eqnarray}
where the subscript c and b denote the cold dark matter and baryons,
respectively, ${\bf V}_{{\rm bc}}$ ($\equiv{\bf V}_{{\rm b}}-{\bf V}_{{\rm c}}$)
is the large-scale streaming velocity of baryons against dark matter,
${\bf k}$ is a given wavenumber, $H$ is the Hubble parameter, $\bar{T}_{\gamma}$
is the average radiation temperature, $k_{B}$ is the Boltzmann constant,
$\mu$ is the mean molecular weight, $m_{{\rm H}}$ is the hydrogen
mass, $t_{\gamma}=1.17\times10^{12}$~years, and $x_{e}(t)$ is the
global ionized fraction\footnote{The original work by TH, for simplicity, ignores $\delta_{T}$ and
assumes a spatially constant sound speed, and works in the baryon-rest frame. Equation (\ref{eq:TH}) instead uses the CDM-rest frame.}. As long as one limits the density environment to that of the mean-density,
CICsASS correctly solves equation (\ref{eq:TH}) and provides initial
conditions that are accurate to the linear order. 

However, in order to include the effect of the density environment
in addition to that of the streaming-velocity environment, equation
(\ref{eq:TH}) is no longer valid and a more accurate description
is required. In addition to ${\bf V}_{{\rm bc}}$, a given large-scale
patch will have in general non-zero overdensity ($\Delta$), velocity
divergence ($\Theta$) and temperature fluctuation ($\Delta_{T}$).
This induces coupling of large-scale and small-scale fluctuations
(denoted ``mode-mode coupling'' hereafter), and was formulated by
Ahn16 into the following perturbation
equation:
\begin{eqnarray}
\frac{\partial\delta_{{\rm c}}}{\partial t} & = & -(1+\Delta_{{\rm c}})\theta_{{\rm c}}-\Theta_{{\rm c}}\delta_{{\rm c}},\nonumber \\
\frac{\partial\theta_{{\rm c}}}{\partial t} & = & -\frac{3}{2}H^{2}\Omega_{m}\left(f_{{\rm c}}\delta_{{\rm c}}+f_{{\rm b}}\delta_{{\rm b}}\right)-2H\theta_{{\rm c}},\nonumber \\
\frac{\partial\delta_{{\rm b}}}{\partial t} & = & -ia^{-1}{\bf V}_{{\rm bc}}\cdot{\bf k}\delta_{{\rm b}}-(1+\Delta_{{\rm b}})\theta_{{\rm b}}-\Theta_{{\rm b}}\delta_{{\rm b}},\nonumber \\
\frac{\partial\theta_{{\rm b}}}{\partial t} & = & -ia^{-1}{\bf V}_{{\rm bc}}\cdot{\bf k}\theta_{{\rm b}}-\frac{3}{2}H^{2}\Omega_{m}\left(f_{{\rm c}}\delta_{{\rm c}}+f_{{\rm b}}\delta_{{\rm b}}\right)-2H\theta_{{\rm b}}\nonumber \\
&&+a^{-2}\frac{k_{B}\bar{T}}{\mu m_{{\rm H}}}k^{2}\left\{ \left(1+\Delta_{{\rm b}}\right)\delta_{T}+\left(1+\Delta_{T}\right)\delta_{{\rm b}}\right\} ,\nonumber \\
\frac{\partial\delta_{T}}{\partial t} & = & \frac{2}{3}\left\{ \frac{\partial\delta_{{\rm b}}}{\partial t}+\frac{\partial\Delta_{{\rm b}}}{\partial t}\left(\delta_{T}-\delta_{{\rm b}}\right)+\frac{\partial\delta_{{\rm b}}}{\partial t}\left(\Delta_{T}-\Delta_{{\rm b}}\right)\right\} \nonumber \\
&&-\frac{x_{e}(t)}{t_{\gamma}}a^{-4}\frac{\bar{T}_{\gamma}}{\bar{T}}\delta_{T}.
\label{eq:Ahn}
\end{eqnarray}
Equation (\ref{eq:Ahn}) is the governing equation for the evolution
of small-scale perturbations ($\delta_{{\rm c}}$, $\delta_{{\rm b}}$,
$\theta_{{\rm c}}$, $\theta_{{\rm b}}$, $\delta_{T}$) under the
influence of local large-scale environment ($\Delta_{{\rm c}}$, $\Delta_{{\rm b}}$,
$\Theta_{{\rm c}}$, $\Theta_{{\rm b}}$, $\Delta_{T}$), accurate
to the leading-order. Equation (\ref{eq:Ahn}) is developed under the CDM-rest frame, such
that an observer is moving at the bulk velocity of CDM particles, ${\bf V}_{\rm c}$, inside the patch.
This equation is based on the assumption that
the two scales are well separated, and thus this formalism is a peak-background
split scheme. We consider the variation of the large-scale fluctuations
at a scale of 4 comoving Mpc, which is a natural choice because the
streaming-velocity at recombination is coherent at a few comoving
Mpc (TH).

We introduce a newly developed, cosmological initial condition generator
BCCOMICS (Baryon-Cold dark matter COsMological Initial Condition generator
for Small-scales) that incorporates the full impact of large-scale
environment on small-scale fluctuations. Obviously, we developed BCCOMICS
because there has not been any initial condition generator that fully
incorporates this effect. The large-scale environments are realized
as a set of three-dimensional fields and their evolution is calculated
properly (Section \ref{subsec:large_scale}). BCCOMICS then solves
the perturbation equation (Equ. \ref{eq:Ahn}) and generates three-dimensional
fields of small-scale quantities (Section \ref{subsec:small_scale}).
Incorporating non-zero overdensity environment requires a careful
reassignment of cosmological parameters and scaling laws, which will
be described in Section \ref{sec:simulating_peaks_voids} in detail. 

\subsection{Evolution of large-scale variance\label{subsec:large_scale}}

We need to understand how large-scale ($\gtrsim$ a few Mpc) fluctuations
vary in space, and how they evolve in time until a dedicated epoch
for the initial condition is reached. 
This is because the impact of the environment
is continuous in time as seen in equation (\ref{eq:Ahn}). For any
given patch chosen out of many large-scale patches with varying physical
properties, once we are able to track its evolution, we obtain the
full understanding of a given environment and this can be fed into
equation (\ref{eq:Ahn}). This evolution, if in the linear regime,
can be easily obtained by Boltzmann solvers such as CAMB, which solves
Eulerian linear evolution equations of all components in the standard
$\Lambda$CDM cosmology. Nevertheless, it would be somewhat more efficient
computationally (we will justify this below) and also intuitive to
have an approximation based on a simplified evolution equation.

Indeed, the evolution of large-scale variables after recombination
can be well approximated by solving for the growth of 4 independent
modes under a very simplified evolution equation (Ahn16) given by
\begin{eqnarray}
\frac{\partial\Delta_{+}}{\partial t} & = & -\Theta_{+},\nonumber \\
\frac{\partial\Theta_{+}}{\partial t} & = & -\frac{3}{2}H^{2}\Omega_{m}\Delta_{+}-2H\Theta_{+},\nonumber \\
\frac{\partial\Delta_{-}}{\partial t} & = & -\Theta_{-},\nonumber \\
\frac{\partial\Theta_{-}}{\partial t} & = & -2H\Theta_{-},\label{eq:background_easy}
\end{eqnarray}
where $\Delta_{+}\equiv f_{{\rm c}}\Delta_{{\rm c}}+f_{{\rm b}}\Delta_{{\rm b}}$,
$\Theta_{+}\equiv f_{{\rm c}}\Theta_{{\rm c}}+f_{{\rm b}}\Theta_{{\rm b}}$,
$\Delta_{-}\equiv\Delta_{{\rm c}}-\Delta_{{\rm b}}$, and $\Theta_{-}\equiv\Theta_{{\rm c}}-\Theta_{{\rm b}}$
with $f_{{\rm c}}\equiv\bar{\rho}_{{\rm c}}/(\bar{\rho}_{{\rm c}}+\bar{\rho}_{{\rm b}})$
and $f_{{\rm b}}\equiv1-f_{{\rm c}}$. They are growing, decaying,
compensated, and streaming modes, which are solutions to these linear
equations (equation \ref{eq:background_easy}). The growing and decaying
modes compose the adiabatic perturbation, and the compensated and
streaming modes compose the isocurvature perturbation. Note that the
isocurvature perturbation can be sourced primordially by the cosmic
inflation in some inflation models, and secularly from BAO
during the recombination epoch (\citealt{Tseliakhovich2010,Barkana2011}). 
We only consider the latter possibility, which is self-consistently
calculated by CAMB.

Equation (\ref{eq:background_easy}) is in this simple form because
(1) fluctuation in the radiation component quickly decays in time
after recombination, (2) the large-scale ($\gtrsim$ a few Mpc) variance
is likely to be free from the mode-mode coupling and (3) the pressure
term $a^{-2}\frac{k_{B}\bar{T}}{\mu m_{{\rm H}}}k^{2}\left\{ \Delta_{T}+\Delta_{{\rm b}}\right\} $
is negligible due to smallness of $k$. Then the temporal evolution
of these variables for a given patch is given by a linear combination
of these modes. Finally, we evolve $\Delta_{T}$ passively, after
solving for the evolution of \{$\Delta_{{\rm c}}$, $\Delta_{{\rm b}}$,
$\Theta_{{\rm c}}$, $\Theta_{{\rm b}}$\} beforehand, by integrating
the following equation (identical to equation 8 of \citealt{Naoz2005}):
\begin{equation}
\frac{\partial\Delta_{T}}{\partial t}=\frac{2}{3}\frac{\partial\Delta_{{\rm b}}}{\partial t}+\frac{x_{e}(t)}{t_{\gamma}}a^{-4}\left\{ \Delta_{\gamma}\left(\frac{5\bar{T}_{\gamma}}{4\bar{T}}-1\right)-\Delta_{T}\frac{\bar{T}_{\gamma}}{\bar{T}}\right\} ,\label{eq:DT_evol}
\end{equation}
where we use the output transfer function from CAMB for the radiation
fluctuation $\Delta_{\gamma}(z)$, and the initial temperature fluctuation
$\Delta_{T}(z_{i})$ is fixed by the scheme by \citet{Naoz2005}:
\begin{equation}
\Delta_{T}(z_{i})=\Delta_{T_{\gamma}}\left(5-\frac{4\bar{T}}{\bar{T}_{\gamma}}\right)+\frac{t_{\gamma}}{x_{e}}a^{4}\left(\frac{2}{3}\frac{\partial\Delta_{{\rm b}}}{\partial t}-\frac{\partial\Delta_{T_{\gamma}}}{\partial t}\right),\label{eq:DT_ini}
\end{equation}
where all the terms of the right hand side are evaluated at $z_{i}$.
Note that equations (\ref{eq:background_easy}-\ref{eq:DT_ini}) can
be all solved in ${\bf k}$-space, and an ${\bf r}$-space map can
be obtained from ${\bf k}$-space quantity by randomization (Equ.
12 of Ahn16) and Fourier transformation.

We note that it is important to consider all the 4 modes in evolving
equation (\ref{eq:background_easy}), because the motion of baryons
are different from that of CDM particles. This difference in motion
among the two fluid components results in non-zero values of the compensated
and streaming modes. These modes even dominate over the growing and
decaying modes around the recombination epoch (see Fig. 9 of Ahn16).
Therefore, it is a bad practice to use only the matter component (growing
and decaying modes), or even only the growing mode as is done in some
structure formation simulations. At the same time, one should consider
the non-negligible radiation component ($\Omega_{r}\equiv\rho_{r}/\rho_{{\rm crit}}$)
at least in the global evolution of $\Omega_{m}$, which makes $\Omega_{m}\ne1$
for quite long after recombination. Cosmological N-body+hydrodynamics
codes such as Enzo and Gadget used not to include the radiation component.
For Enzo at least, therefore, we have now implemented non-zero $\Omega_{r}$
and Enzo correctly calculates the corresponding $\Omega_{m}(z)$ and the
Hubble parameter $H(z)$\footnote{The up-to-date development version of Enzo, which is downloadable
from http://enzo-project.org, reflects this implementation.}. Otherwise, a spurious effect in structure formation simulation will
occur if one applies the transfer function from CAMB, which of course
considers the radiation component, to N-body+hydro codes which
do not have the radiation component.

In practice, the following steps are performed for calculating the
evolution of a patch. We refer readers to section 2.2 and Appendix
of Ahn16 for details. First, at $z_{{\rm re}}$, we use the transfer
function from a widely-used Boltzmann solver CAMB. By convolving the
transfer function with a Gaussian random seed, 3D maps of \{$\Delta_{{\rm c}}$,
$\Delta_{{\rm b}}$, $\Theta_{{\rm c}}$, $\Theta_{{\rm b}}$, ${\bf V}_{{\rm c}}$,
${\bf V}_{{\rm b}}$\}, or \{$\Delta_{+}$, $\Delta_{-}$, $\Theta_{+}$,
$\Theta_{-}$, ${\bf V}_{{\rm c}}$, ${\bf V}_{{\rm b}}$\} at $z_{{\rm re}}$
can be realized. Second, we numerically solve for the temporal evolutions
of \{$\Delta_{+}$, $\Delta_{-}$, $\Theta_{+}$, $\Theta_{-}$, ${\bf V}_{{\rm bc}}$\}.
Evolution of each of these mode variables is fully described by a
single corresponding numerical solution (let's say $\mathcal{F}(z)$),
because equation (\ref{eq:background_easy}) is linear. Third, at
any $z$, one simply multiplies the fluctuation value at $z_{{\rm re}}$
to $\mathcal{F}(z)$ and obtain the evolved value at $z$. For example,
if $\mathcal{F}_{\Delta_{+}}(z)$ is the solution of $\Delta_{+}$
with normalization convention $\mathcal{F}_{\Delta_{+}}(z_{{\rm re}})=1$,
$\Delta_{+}(z)=\Delta_{+}(z_{{\rm re}})\mathcal{F}_{\Delta_{+}}(z)$.
Fourth, as for $\Delta_{T}$, one can use either the actual evolution
(obtained by integrating equation \ref{eq:DT_evol}) or a fitting
formula given by equation (30) of Ahn16: thanks to the quick coupling
of $\Delta_{T}$ to $\Delta_{{\rm b}}$ after $z\simeq500$ the fitting
formula is given in terms of $\Delta_{{\rm b}}$, and smallness of
$\Delta_{T}$ ($\lesssim10^{-5}$) before $z\simeq500$ allows us
to simply set $\Delta_{T}=0$ when $z\gtrsim500$. Then these solutions
are fed in equation (\ref{eq:Ahn}) when solving for the evolution
of small-scale fluctuations for any given patch. 

\begin{figure*}
\gridline{
\fig{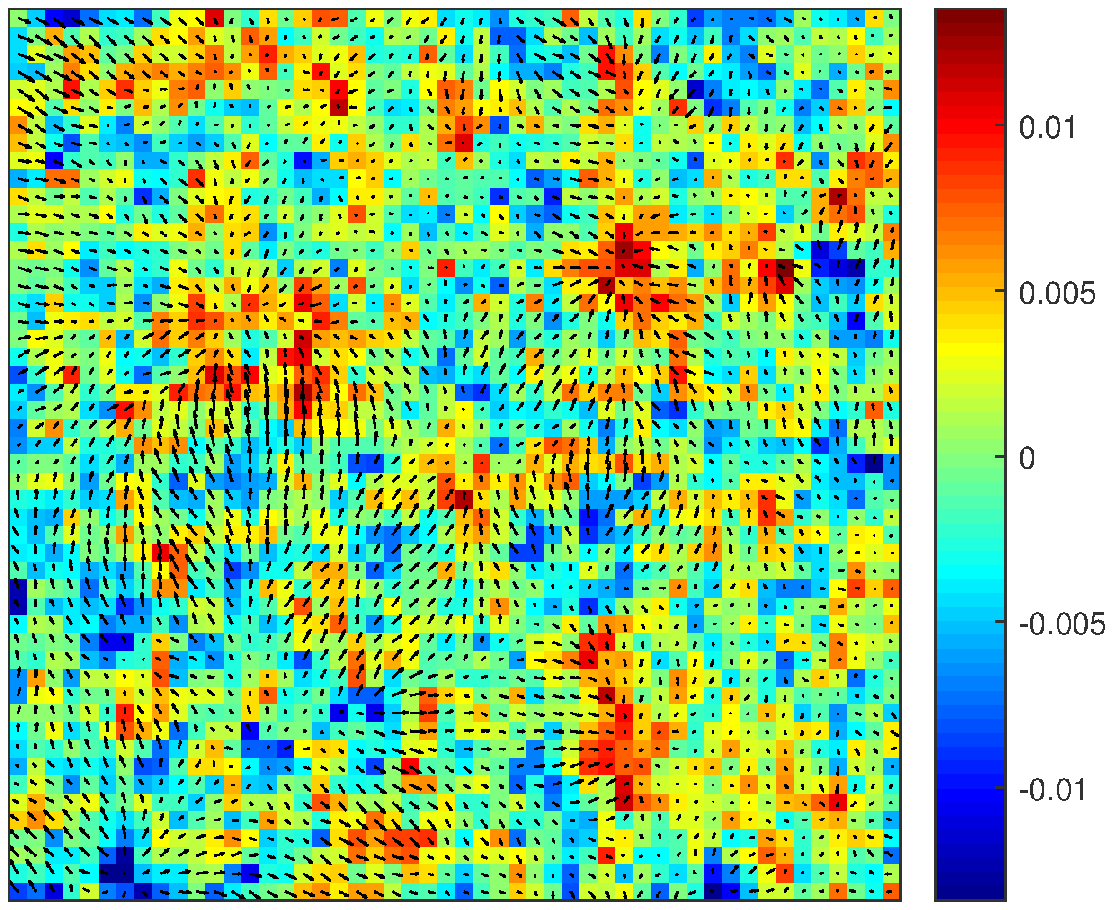}{0.34\textwidth}{(a)}
\fig{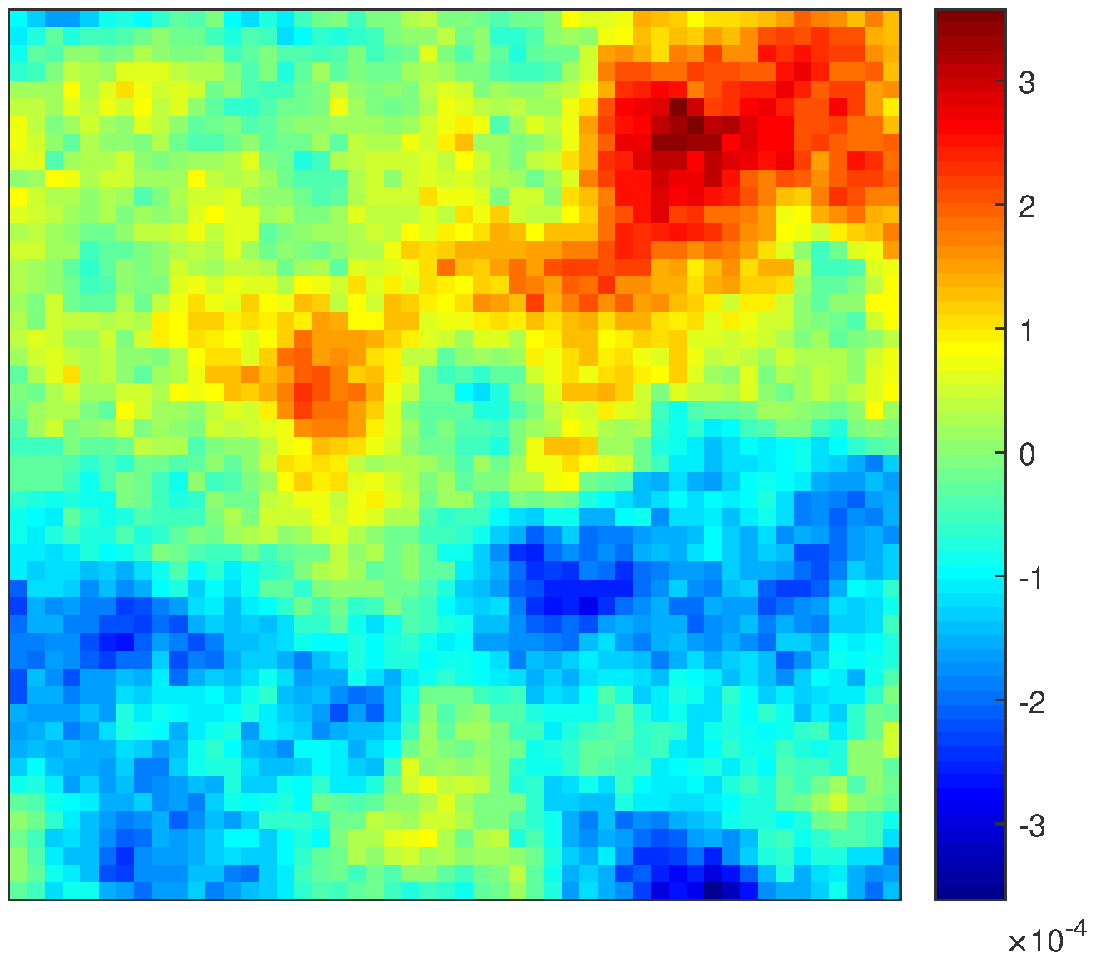}{0.33\textwidth}{(b)}
\fig{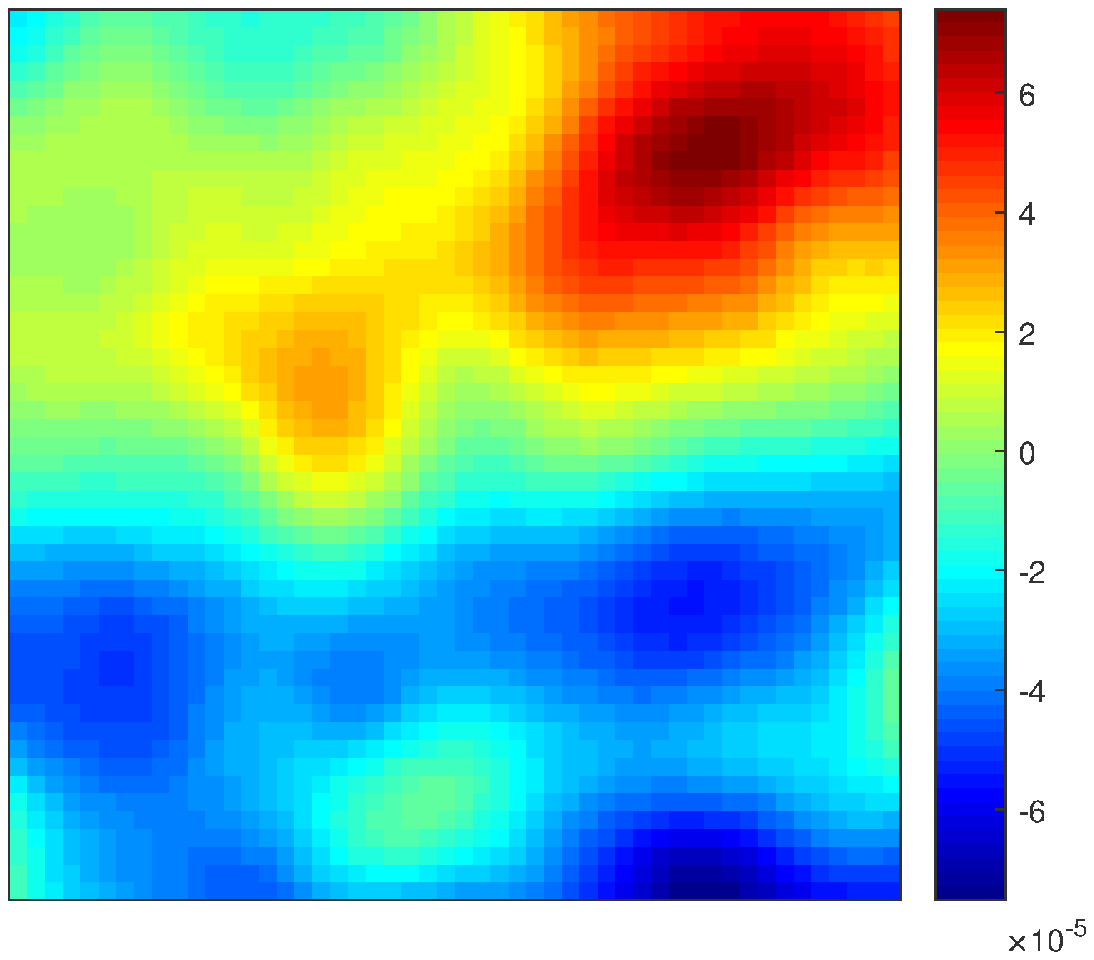}{0.33\textwidth}{(c)}
}

\gridline{
\fig{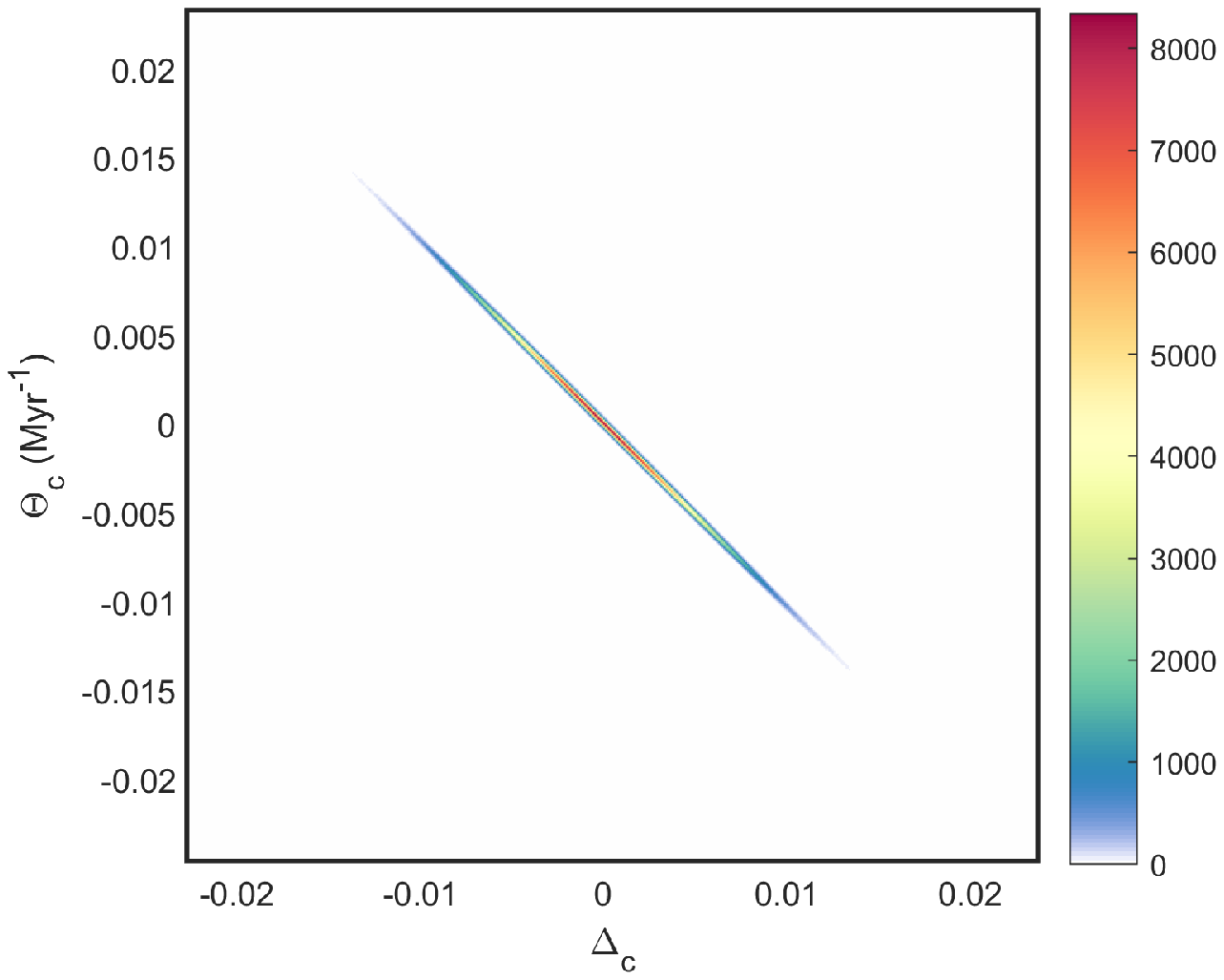}{0.33\textwidth}{(d)}
\fig{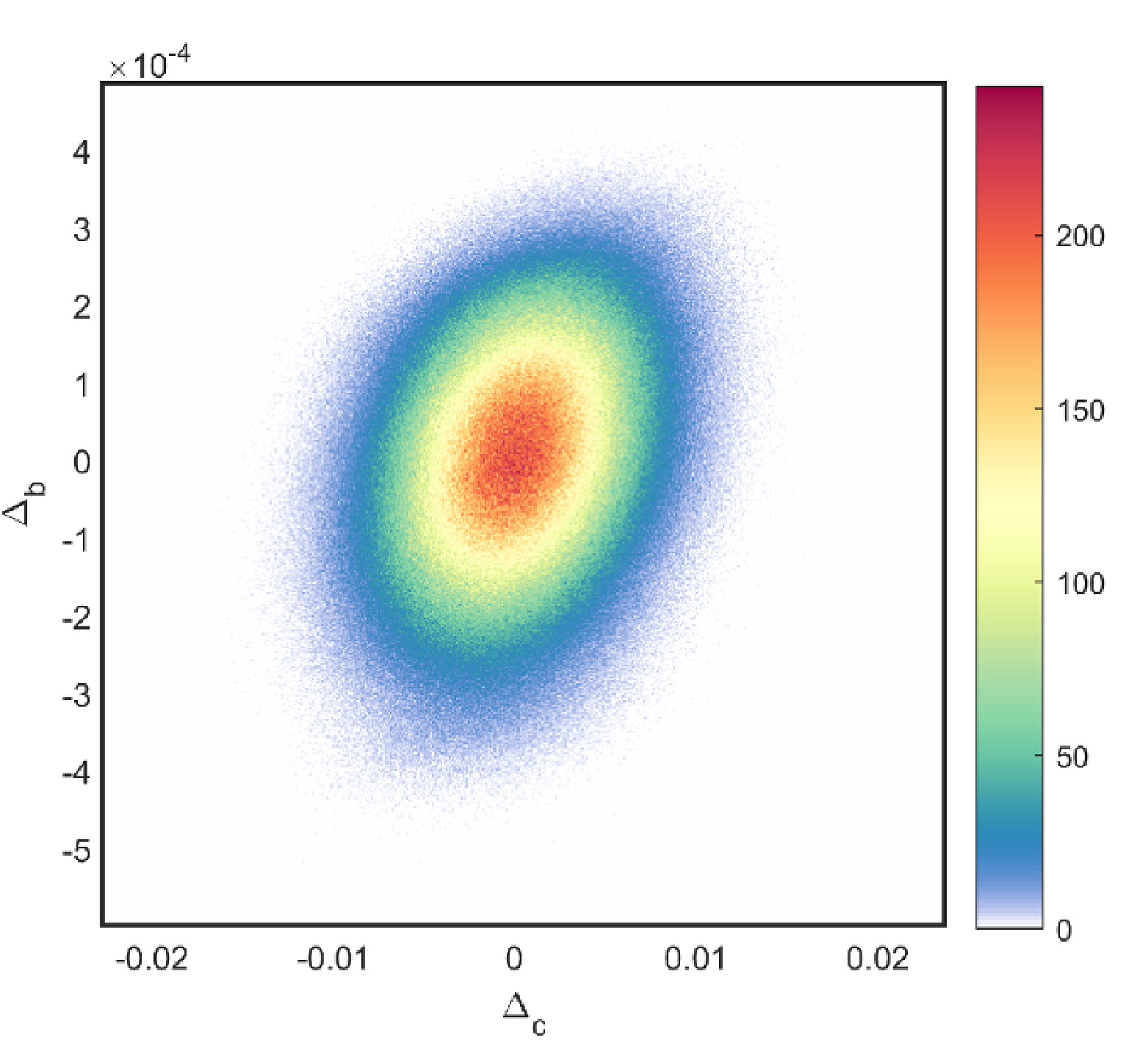}{0.33\textwidth}{(e)}
\fig{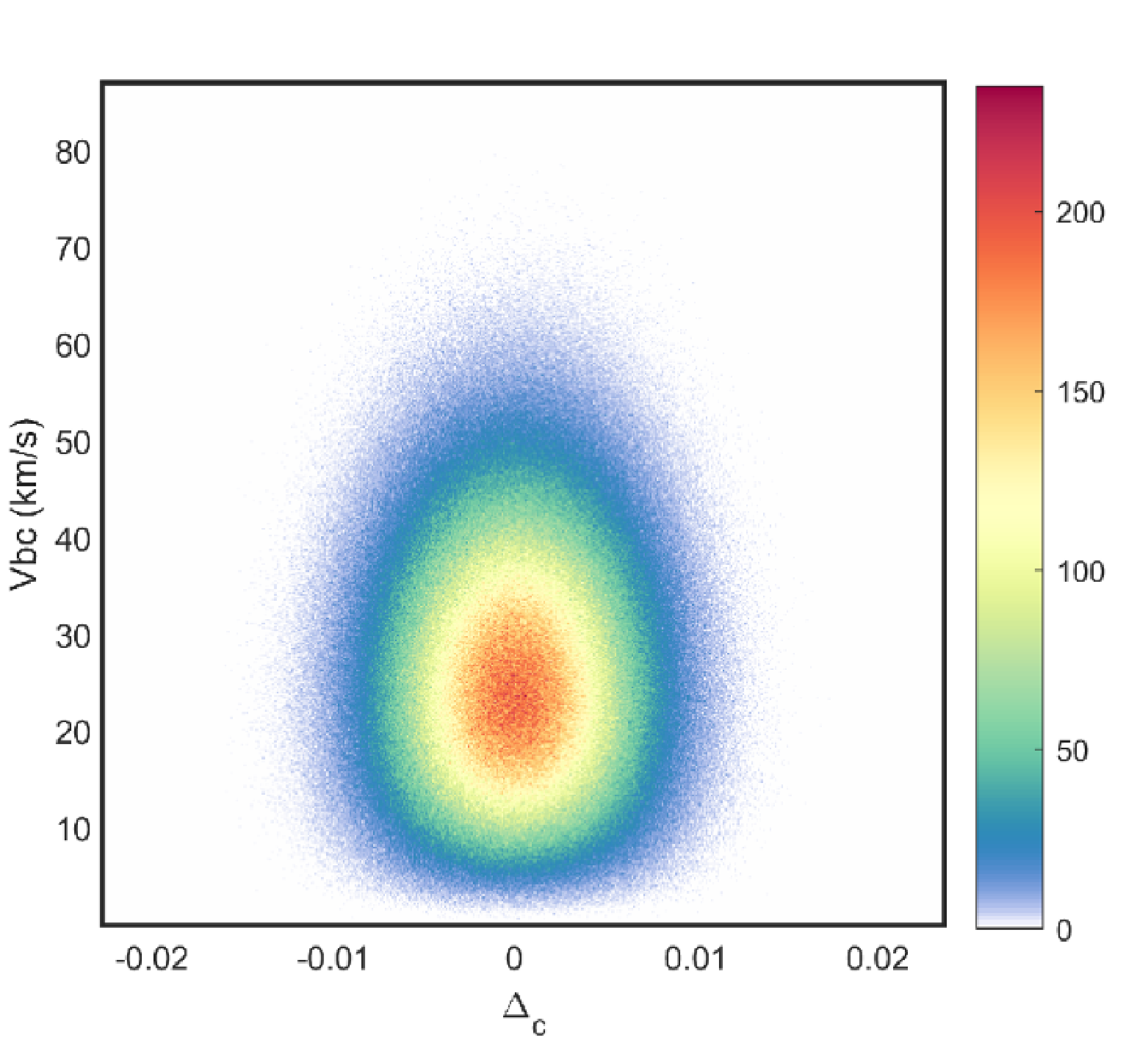}{0.33\textwidth}{(f)}
}

\caption{(a) Map of CDM overdensity $\Delta_{{\rm c}}$ (colored cells) and
the streaming velocity ${\bf V}_{{\rm cb}}\equiv{\bf V}_{{\rm c}}-{\bf V}_{{\rm b}}$
(arrows) on a slice of $200^{2}\,{\rm Mpc^{2}}$ are containing $50^{2}$
cells. This is an arbitrarily chosen part of the periodic box with
an actual volume of $604^{3}\,{\rm Mpc}^{3}$. (b) Map of baryon overdensity
$\Delta_{{\rm b}}$ on the same slice. (c) Map of temperature overdensity
$\Delta_{T}$ on the same slice. (d) Distribution of $\Delta_{{\rm c}}$
and the CDM velocity divergence $\Theta_{{\rm c}}$, with colors representing
the number of cells in sampling bins. (e) Distribution of $\Delta_{{\rm c}}$
and $\Delta_{{\rm b}}$. (f) Distribution of
$\Delta_{{\rm c}}$ and $V_{{\rm bc}}\equiv\left|{\bf V}_{{\rm cb}}\right|$.
All panels are based on quantities at $z=1000$. \label{fig:z1000}}
\end{figure*}

\begin{figure*}
\gridline{
\fig{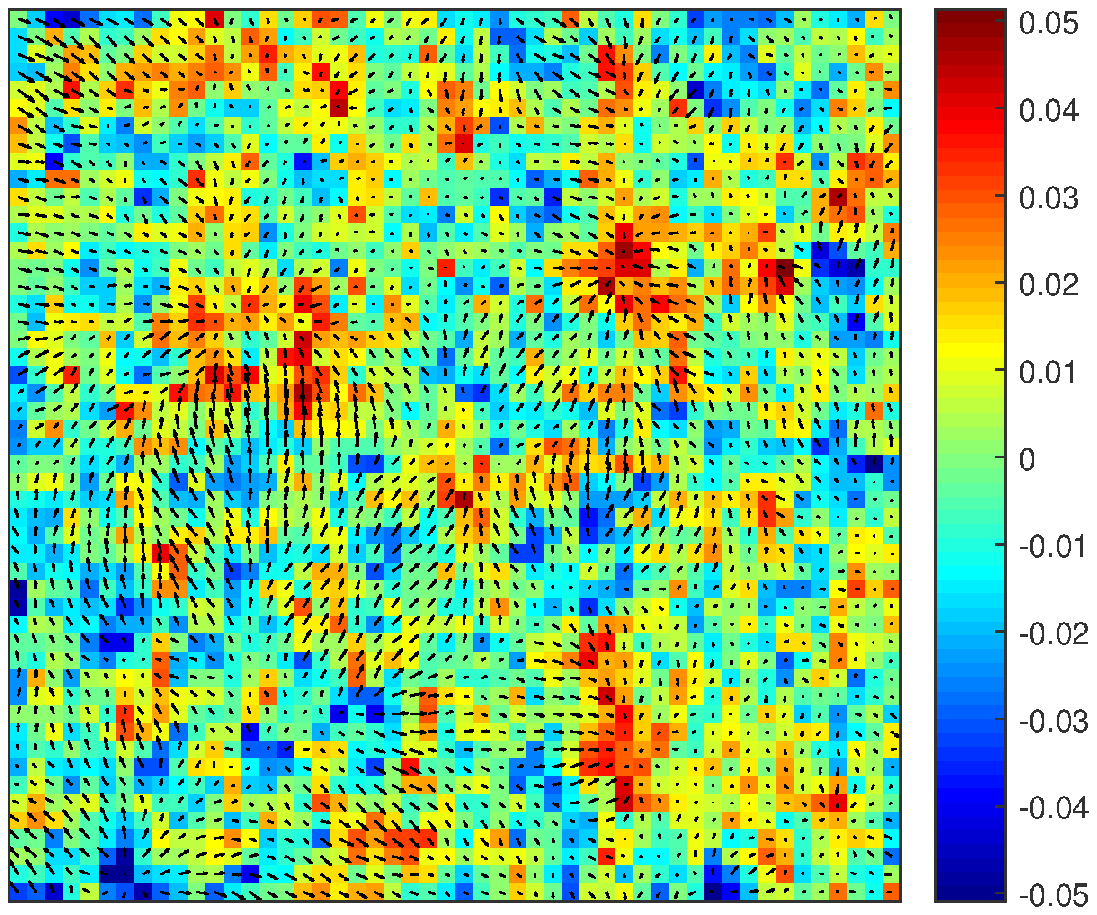}{0.33\textwidth}{(a)}
\fig{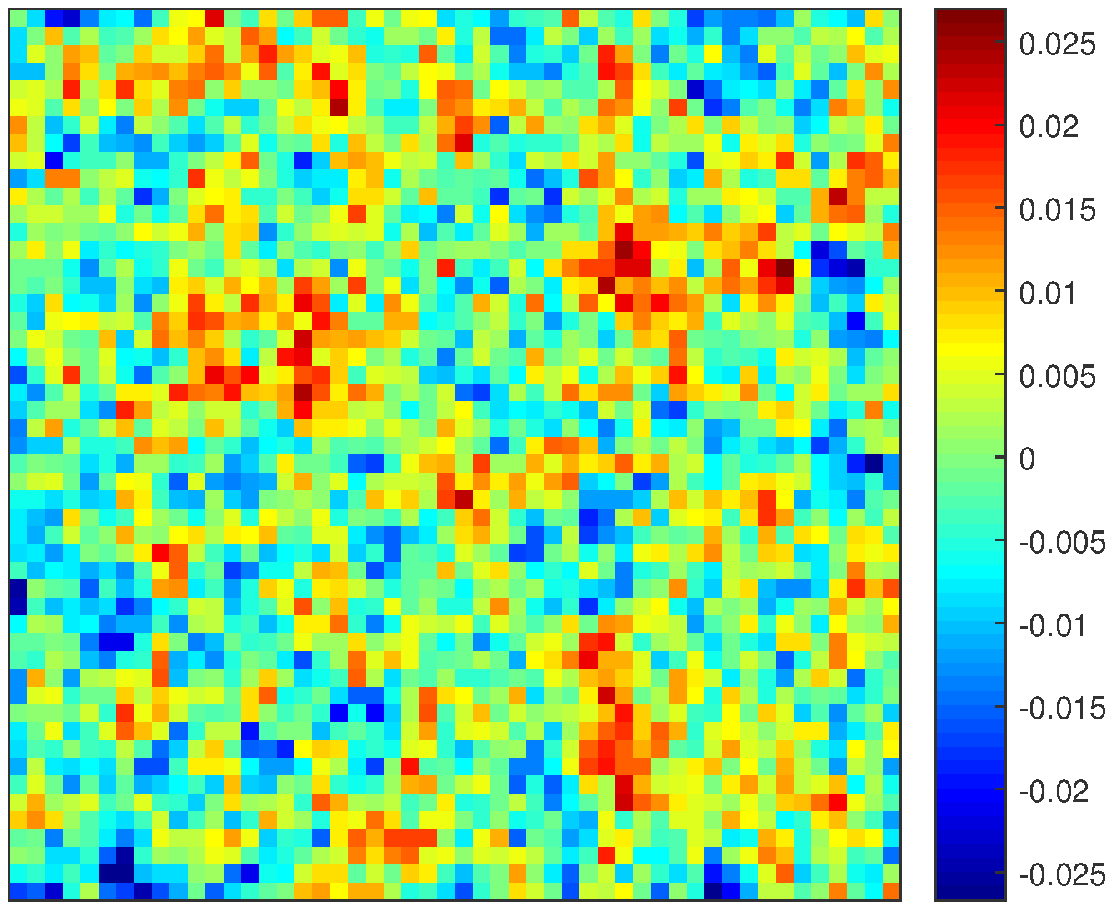}{0.337\textwidth}{(b)}
\fig{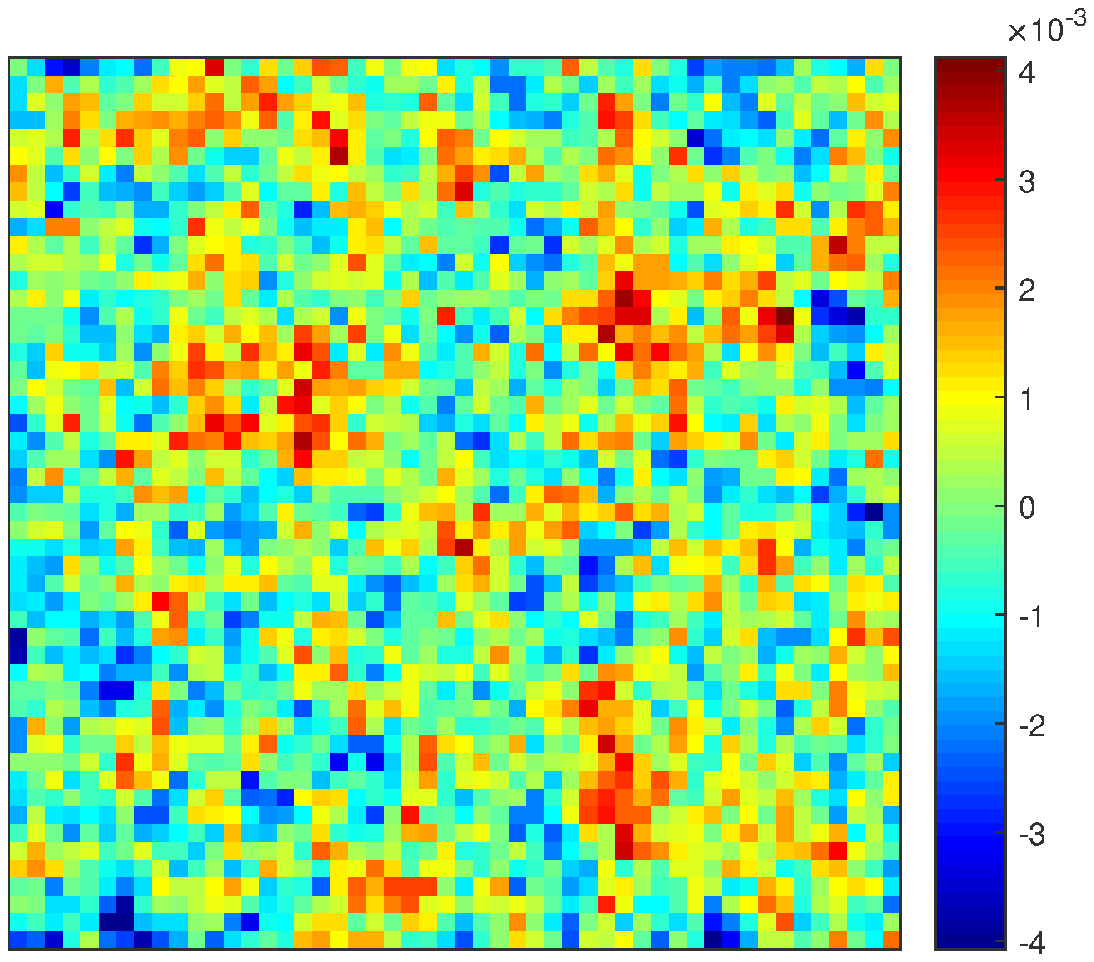}{0.328\textwidth}{(c)}
}

\gridline{
\fig{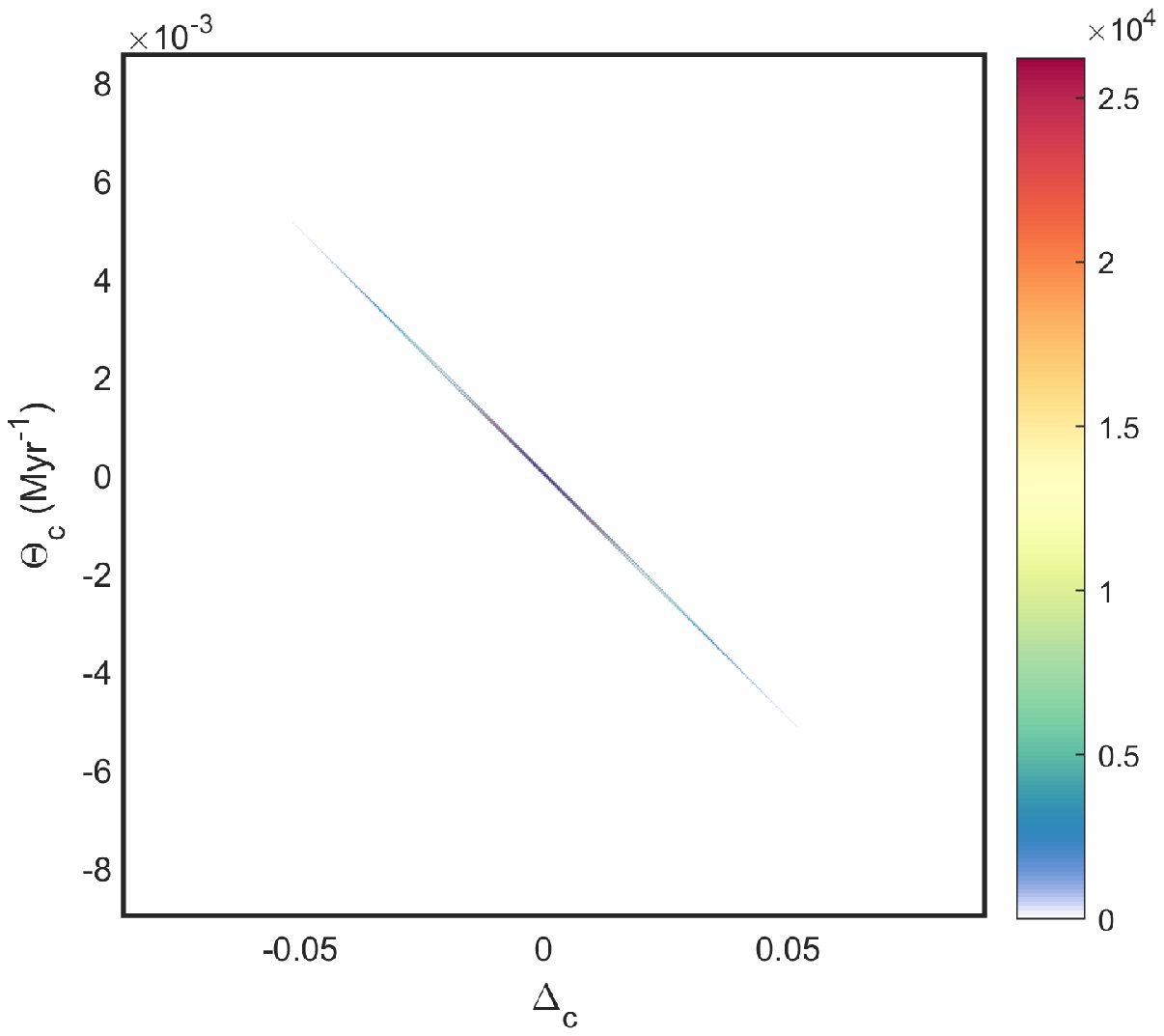}{0.313\textwidth}{(d)}
\fig{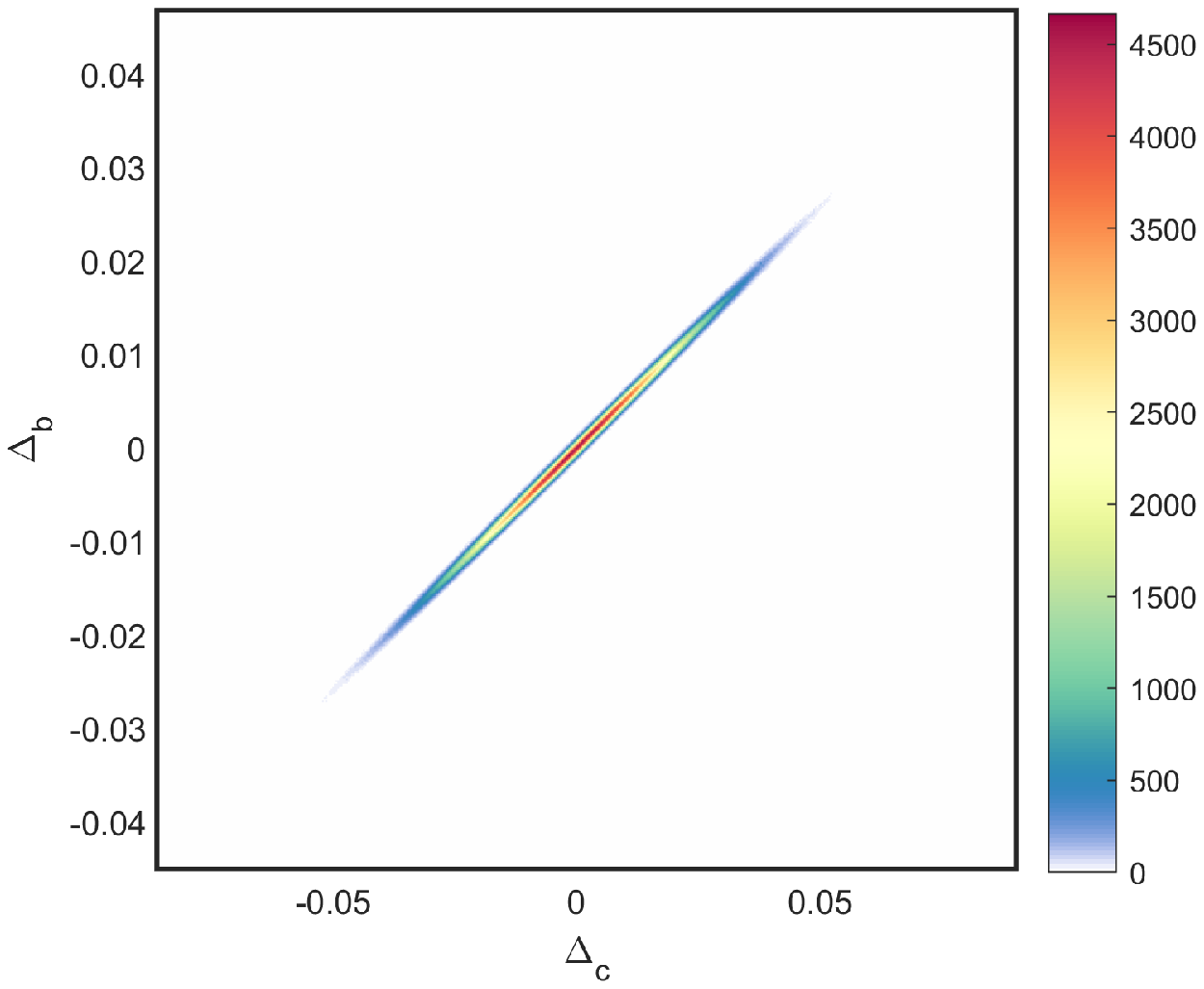}{0.313\textwidth}{(e)}
\fig{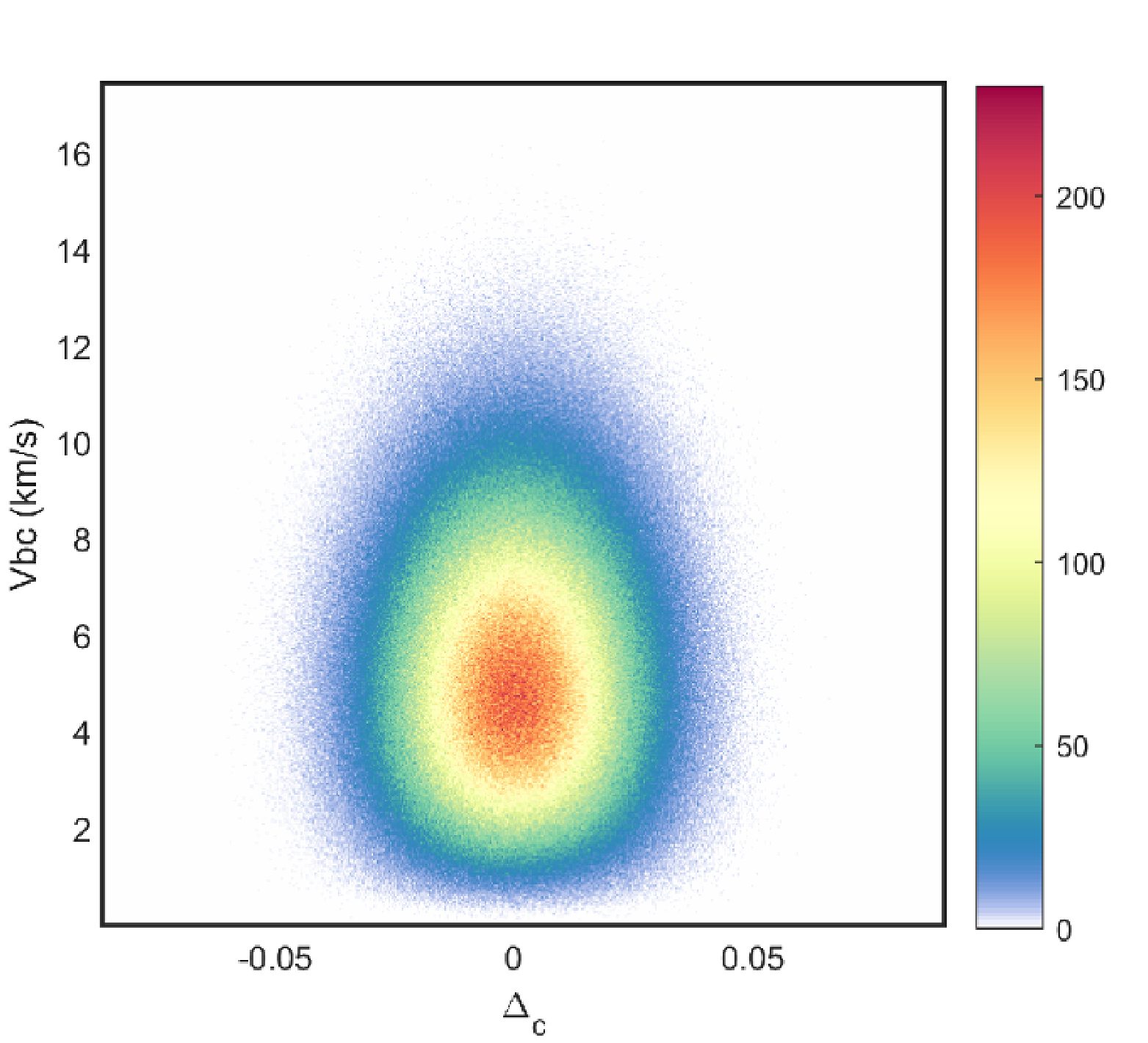}{0.313\textwidth}{(f)}
}

\caption{Same as Figure \ref{fig:z1000}, except that all panels are now based
on quantities at $z=200$. \label{fig:z200}}
\end{figure*}

An important physical intuition can be seen by comparing spatial maps
and correlations of these variables at $z\simeq z_{{\rm re}}$ and
a target redshift, e.g., $z_{i}=200$. Figure \ref{fig:z1000} and
Figure \ref{fig:z200} show maps and histograms at $z=1000$ and $z=200$,
respectively. As seen in Figure \ref{fig:z1000}(a), ${\bf V}_{{\rm cb}}\equiv{\bf V}_{{\rm c}}-{\bf V}_{{\rm b}}$
(arrows) is dominated by ${\bf V}_{{\rm c}}$ and thus show convergence
into peaks and divergence from voids. Comparing Figure \ref{fig:z1000}(a)
and \ref{fig:z1000}(b), we find that at $z=1000$ $\Delta_{{\rm c}}$
dominates over $\Delta_{{\rm b}}$ in amplitude, and these two are
very weakly correlated, as also seen in Fig. \ref{fig:z1000}(e).
This is due to the tight coupling of baryons to photons during recombination,
which can also be seen in the map of $\Delta_{T}$ (Fig. \ref{fig:z1000}c)
showing the characteristic feature of the baryon acoustic oscillation
(BAO). CDM moves mainly through its self-gravity, producing a very
tight correlation between $\Delta_{{\rm c}}$ and $\Theta_{{\rm c}}$
(Fig. \ref{fig:z1000}d). $V_{{\rm bc}}$ and $\Delta_{{\rm c}}$
are not correlated at all (Fig. \ref{fig:z1000}f).

\begin{figure}
\includegraphics[width=0.4\textwidth]{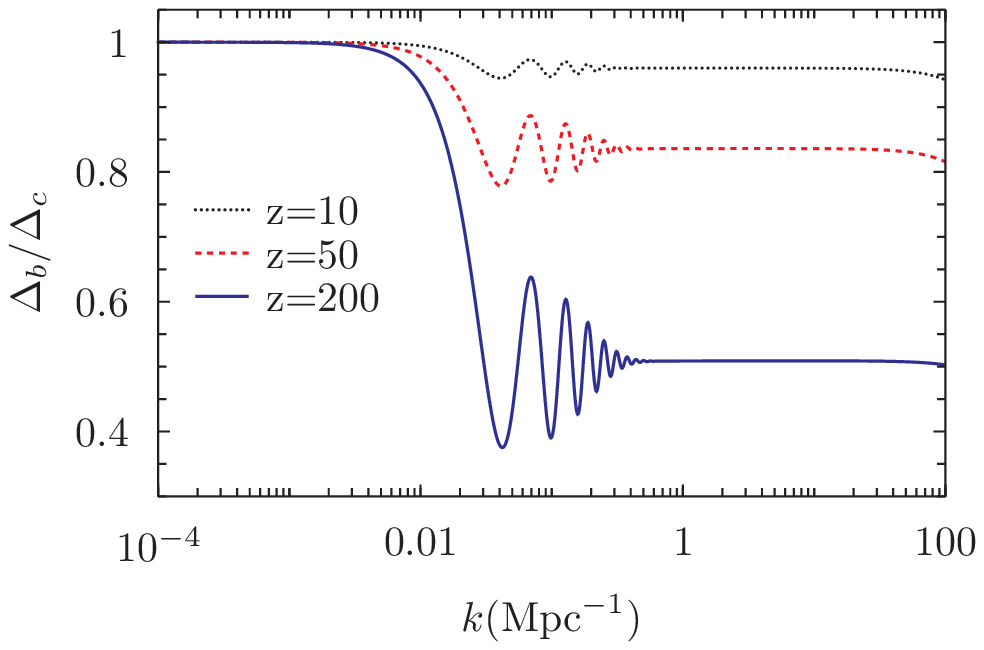}

\caption{Overdensity ratio $\Delta_{{\rm b}}/\Delta_{{\rm c}}$ calculated
by the linear Boltzmann solver at different wavenumbers $k$ and redshifts
$z$.\label{fig:bar2cdm}}

\end{figure}

The main difference between the two epochs lies in the behavior of
baryon fluctuations. At $z=200$, aside from the amplitude, maps of $\Delta_{{\rm c}}$,
$\Delta_{{\rm b}}$ and $\Delta_{T}$ are almost indistinguishable
(Figs. \ref{fig:z200}a-\ref{fig:z200}c). This is because the baryon
fluctuation is now governed predominantly by gravity rather than by
photon fluctuation. The correlation between $\Delta_{{\rm c}}$ and
$\Theta_{{\rm c}}$ is even tighter (Fig. \ref{fig:z200}d) than it
is at $z=1000$, and now $\Delta_{{\rm b}}$ is strongly correlated
with $\Delta_{{\rm c}}$ (Fig. \ref{fig:z200}e). The latter fact
is in stark contrast with the loose correlation seen at $z=1000$
(Fig. \ref{fig:z1000}e). Because of this, one may assume that baryons
are ``locked'' into CDM such that $\rho_{{\rm b}}/\rho_{{\rm c}}=\Omega_{{\rm b}}/\Omega_{{\rm c}}$
in all patches, and consider $\Delta_{{\rm c}}$ as the only important
density environment after $z=200$. This is indeed an approximation
appropriate for the study of first-galaxy formation, and allows one
to use a periodic boundary condition without worrying the net inflow
of baryons to the simulation box. We therefore take this approximation
and isolate the simulation box with baryon and CDM contents fixed
in this paper. However, we note that the full degree of such a ``locking''
has not happened yet, because $\Delta_{{\rm b}}/\text{\ensuremath{\Delta}}_{{\rm c}}\ne1$
but $\Delta_{{\rm b}}/\text{\ensuremath{\Delta}}_{{\rm c}}\simeq0.5$.
This makes $\rho_{{\rm b}}/\rho_{{\rm c}}$ slightly off from the
cosmic abundance $\Omega_{{\rm b}}/\Omega_{{\rm c}}$, even though
the smallness of $\Delta_{{\rm b}}$ and $\text{\ensuremath{\Delta}}_{{\rm c}}$
at $z=200$ make this approximation acceptable. Nevertheless, in time,
the value of $\Delta_{{\rm b}}/\text{\ensuremath{\Delta}}_{{\rm c}}$
gradually approaches $1$, but still with some variance over the large
scale ($k\lesssim0.1\,{\rm Mpc^{-1}}$) (Fig. \ref{fig:bar2cdm}).
This is indeed a result of BAO. For a precision cosmology with galaxies
or the intergalactic medium (IGM), for example, one should therefore consider other large-scale
variables as well as $\Delta_{{\rm c}}$. $V_{{\rm bc}}$ and $\Delta_{{\rm c}}$
still remain uncorrelated (Fig. \ref{fig:z200}f). 

\subsection{Anisotropic transfer function and real-space fluctuations\label{subsec:small_scale}}

All the transfer functions (let us denote it by $\mathcal{T}_{s}({\bf
k},\,z;\,{\bf V}_{\rm bc},\,\{\Delta_{L}\})$\footnote{$\mathcal{T}_{l}$ depends on several
  environmental variables, in 
principle. Nevertheless, strong correlation at $z=1000$ between $\Delta_{{\rm c}}$
and $\Theta_{{\rm c}}$ allows us to remove the dependence on $\Theta_{{\rm c}}$.
In addition, we believe that the dominant effect on small-scale perturbations
comes mainly from $V_{{\rm bc}}$ and $\Delta_{{\rm c}}$ after the
motion of baryons are approximately synchronized with that of CDM.
Therefore, in this paper we do not investigate the dependence on other
variables, $\Delta_{{\rm b}}$, $\Theta_{{\rm b}}$ and $\Delta_{T}$,
in our simple case studies (Section \ref{sec:Application}).} for
small-scale component $s$ under the influence of large-scale 
component $L$; $\left\{ \Delta_{L}\right\} $ is the abbreviation
for all large-scale fluctuations) of small-scale perturbations ($\{\delta_{s}\}$=\{$\delta_{{\rm c}}$,
$\delta_{{\rm b}}$, $\theta_{{\rm c}}$, $\theta_{{\rm b}}$, $\delta_{T}$\})
are anisotropic due to the impact of ${\bf V}_{{\rm bc}}$, as expected
from equation (\ref{eq:Ahn}). This makes it necessary to calculate
$\mathcal{T}_{s}({\bf k},\,z;\,{\bf V}_{\rm bc},\,\{\Delta_{L}\})$ at
every wavenumber ${\bf k}$. Or, we can use an implied dependence
$\mathcal{T}_{s}({\bf k},\,z;\,{\bf V}_{{\rm bc}},\,\{\Delta_{L}\})=\mathcal{T}_{s}(k,\,\mu,\,z;\,V_{{\rm bc}},\,\{\Delta_{L}\})$,
where $\mu\equiv{\bf k}\cdot{\bf V}_{{\rm bc}}/(kV_{{\rm bc}})$.
In practice, we first integrate equation (\ref{eq:Ahn}) and generate
$\mathcal{T}_{s}(k,\,\mu,\,z;\,V_{{\rm bc}},\,\{\Delta_{L}\})$ with
dozens of logarithmic samples $k$ and a uniformly gridded samples
of $\mu$, for a given patch with a specific set
of large-scale fluctuations $\{\Delta_{L}\}$ and ${\bf V}_{{\rm bc}}$. 
(1) This forms a 2D
table of $\mathcal{T}_{s}$ in terms of $\{k,\,\mu_{m}\}$
for a given patch, where subscript $m$ is the integer index for the discretized
values of $\mu$, or $\mu_{m}$.
(2) Then, for a given ${\bf k}$ and ${\bf V}_{{\rm bc}}$,
we perform a 2D interpolation at the actual set of points $\{k,\,\mu\}$
and obtain $\mathcal{T}_{s}({\bf k},\,z;\,{\bf V}_{{\rm bc}},\,\{\Delta_{L}\})$. 
Once we obtain $\mathcal{T}_{s}({\bf k},\,z;\,{\bf V}_{\rm bc},\,\{\Delta_{L}\})$,
with proper normalization (to fluctuations at $z=1000$),  the
power spectrum of quantity $l$ becomes
$P_{s}({\bf k},\,z;\,{\bf V}_{\rm bc},\,\{\Delta_{L}\})=P_{s}(k,\,z=1000)\mathcal{T}_{s}^{2}({\bf k},\,z;\,{\bf V}_{bc},\,\{\Delta_{L}\})$.
(3) We apply two Gaussian random seeds $G_{1}$ and $G_{2}$ to obtain
real and imaginary parts of the ${\bf k}$-space field, where both
$G_{1}$ and $G_{2}$ have mean 0 and variance 1:
\begin{eqnarray}
{\rm Re}(\delta_{s}({\bf k})) & = & G_{1}N^{3}\left(\frac{P_{s}({\bf k})}{2V_{{\rm box}}}\right)^{1/2}{\rm sign}\left[T_{s}({\bf k})\right],\nonumber \\
{\rm Im}(\delta_{s}({\bf k})) & = & G_{2}N^{3}\left(\frac{P_{s}({\bf k})}{2V_{{\rm box}}}\right)^{1/2}{\rm sign}\left[T_{s}({\bf k})\right].
\end{eqnarray}
(4) Of course, real-space variables are all in real numbers, thus
we use the condition $\delta_{s}^{*}({\bf k})=\delta_{s}(-{\bf k})$ after
filling only 1/2 of the allowed ${\bf k}$-space ($^{*}$ denotes
the complex conjugate). In addition, all
monopole terms (${\bf k}=0$) are assumed zero. Removing monopole
terms allows one to use the usual periodic boundary condition, and
we will explain in Section \ref{sec:simulating_peaks_voids} how this
becomes possible even in the presence of non-zero $\Delta_{{\rm c}}$
and $\Theta_{{\rm c}}$. (5) Finally, we take the Fourier transform
of $\delta_{s}({\bf k})$ and obtain $\delta_{s}({\bf x})$. Vectorization
of the baryon velocity field is performed through the relation ${\bf v}({\bf k})=-(ia{\bf k}/k^{2})\theta({\bf k})$
assuming a curl-free velocity field.

The whole process (1) - (5) is straightforward for generating uniform-grid
quantities. For particle quantities (of CDM particles; of baryon particles 
if smoothed particle hydrodynamics is used), we follow
the usual Lagrangian perturbation theory (LPT), relating the displacement
vector to the Eulerian overdensity \citep{Bouchet1992}. Because we
calculate the Eulerian overdensity based on the linear-order Boltzmann
solver CAMB, we restrict our calculation to the 1st-order Lagrangian
perturbation theory (1LPT). A particle is displaced from its Lagrangian
point ${\bf q}$ to its Eulerian point ${\bf x}$ by the displacement
vector ${\bf \Psi}$, as ${\bf x}={\bf q}+{\bf \Psi}$. To linear
order,
\begin{equation}
\delta({\bf q})=-{\bf \nabla}\cdot{\bf \Psi}({\bf q}),\label{eq:LPT}
\end{equation}
which can be again cast into the ${\bf k}$-space quantities
\begin{equation}
{\bf \Psi}=(i{\bf k}/k^{2})\delta\label{eq:LPTk}
\end{equation}
and allows using $\mathcal{T}_{\delta}({\bf k})$ for $\delta({\bf k})$.
It is easy to generate the real-space displacement field through the
Fourier transformation of ${\bf \Psi}({\bf k})$ if displacing particles
from cubically and uniformly spaced positions; in case of a glass,
an interpolation of this uniform-space $\Psi({\bf q})$ field onto
glassy positions is further required. %
For the particle velocity field, consequently, its generation first
takes the steps for the uniform-grid data, and interpolation of
this uniform-grid velocity field onto the displaced positions are
performed. In practice, however, such interpolation becomes unnecessary
if ${\bf \Psi}$ is very small compared to the length scale $\sim\left|\Psi/\nabla\cdot{\bf \Psi}\right|$.

\section{How to simulate density peaks and voids\label{sec:simulating_peaks_voids}}

Because BCCOMICS solves equation (\ref{eq:Ahn}) which includes possible
couplings between large-scale (in terms of $\left\{ \Delta_{l}\right\} $)
and small-scale modes (in terms of $\left\{ \delta_{s}\right\} $),
BCCOMICS provides so far the most accurate (and only) initial condition
regarding the large-scale environmental effect on small-scale fluctuations.
To take advantage of this fact and simulate structure formation in
overdense and underdense regions, however, we need to take further
steps than those required for a mean-density environment. In this
section, we lay out a strategy for achieving this goal.

\subsection{Local Hubble parameter and the Friedmann equation\label{subsec:localH}}

We will take an overdense (underdense) patch as a separate universe.
There can be other ways of simulating an overdense patch, such as
taking a much larger volume as a simulation box and zooming into a
patch of interest with e.g. nested grids (e.g. \citealt{Oshea2015}). 
Nevertheless, this scheme
has a merit of allowing the periodic boundary condition, because an
overdense (underdense) patch in our universe is now treated as a mean-density
patch in a universe with different cosmological parameters.

Because the patch will detach from the global Hubble flow, almost
all relevant cosmological parameters should be redefined. Anything
redefined in such a separate universe will be called ``local'' and
the relevant symbol will be capped by $\tilde{}$. The mass conservation
of such a patch first defines the local Hubble parameter (\citealt{Goldberg2004}):
\begin{equation}
\tilde{H}=H-\frac{\dot{\Delta}}{3(1+\Delta)},\label{eq:hubble}
\end{equation}
where $\Delta$ is the matter overdensity of a given patch, $\dot{}=\frac{d}{dt}$
with cosmic time $t$, and we assume that the net influxes of CDM
  and baryons are both zero on this patch in the local viewpoint, 
or $\tilde{\Theta}_{\rm c}=\tilde{\Theta}_{\rm b}=0$, in order to
incorporate the usual periodic box condition for
simulation. Nevertheless, because we take a patch in a CDM-rest frame,
there will be a bulk flow of baryons with velocity $\tilde{\bf V}_{\rm b}={\bf
  V}_{\rm bc}$.
Note also that $\tilde{\Delta}_{\rm c}=\tilde{\Delta}_{\rm b}=0$ by 
definition, because we take a patch as a separate mean-density universe.

The local Friedmann equation can be written in various forms:
\begin{equation}
1=\tilde{\Omega}_{m}+\tilde{\Omega}_{r}+\tilde{\Omega}_{\Lambda}+\tilde{\Omega}_{K},
\label{eq:identity}
\end{equation}
\begin{equation}
\tilde{H}=\tilde{H}_{0}\left[\tilde{\Omega}_{m,0}\tilde{a}^{-3}+\tilde{\Omega}_{r,0}\tilde{a}^{-4}+\tilde{\Omega}_{\Lambda,0}+\tilde{\Omega}_{K,0}\tilde{a}^{-2}\right]^{1/2},
\label{eq:h_h0}
\end{equation}
\begin{eqnarray}
\tilde{H}=\tilde{H}_{i}&&\left[\tilde{\Omega}_{m,i}\left(\frac{\tilde{a}}{\tilde{a}_{i}}\right)^{-3}+\tilde{\Omega}_{r,i}\left(\frac{\tilde{a}}{\tilde{a}_{i}}\right)^{-4}+\tilde{\Omega}_{\Lambda,i}\right. \nonumber\\
&&\left. +\tilde{\Omega}_{K,i}\left(\frac{\tilde{a}}{\tilde{a}_{i}}\right)^{-2}\right]^{1/2},
\label{eq:h_hi}
\end{eqnarray}
where $\tilde{a}_{0}=1$ convention is used in equation (\ref{eq:h_h0}),
and the subscript $i$ refers to an initial time with the initial
scale factor $\tilde{a}_{i}$. Equation (\ref{eq:hubble}) gives
\begin{equation}
\frac{\tilde{H}_{i}}{H_{i}}=1-\frac{(\dot{\Delta})_{i}}{3H_{i}(1+\Delta_{i})}.
\label{eq:hubble_i}
\end{equation}

The initial local cosmological parameters become
\begin{eqnarray}
\tilde{\Omega}_{m,i}&=&\frac{\tilde{\rho}_{m,i}}{\tilde{\rho}_{{\rm crit},i}}=\frac{(1+\Delta_{i})\rho_{m,i}}{\tilde{\rho}_{{\rm crit},i}}=\frac{(1+\Delta_{i})\rho_{m,i}}{\rho_{{\rm crit},i}}\frac{\rho_{{\rm crit},i}}{\tilde{\rho}_{{\rm crit},i}}\nonumber\\
&=&(1+\Delta_{i})\Omega_{m,i}\left(\frac{\tilde{H}_{i}}{H_{i}}\right)^{-2},
\label{eq:Om_i}
\end{eqnarray}
\begin{equation}
\tilde{\Omega}_{r,i}=\frac{\tilde{\rho}_{r,i}}{\tilde{\rho}_{{\rm crit},i}}=\frac{\rho_{r,i}}{\tilde{\rho}_{{\rm crit},i}}=\Omega_{r,i}\left(\frac{\tilde{H}_{i}}{H_{i}}\right)^{-2},
\label{eq:Or_i}
\end{equation}
\begin{equation}
\tilde{\Omega}_{\Lambda,i}=\frac{\tilde{\rho}_{\Lambda,i}}{\tilde{\rho}_{{\rm crit},i}}=\frac{\rho_{\Lambda,i}}{\tilde{\rho}_{{\rm crit},i}}=\Omega_{\Lambda,i}\left(\frac{\tilde{H}_{i}}{H_{i}}\right)^{-2},
\label{eq:OL_i}
\end{equation}
\begin{equation}
\tilde{\Omega}_{K,i}=1-\left(\tilde{\Omega}_{m,i}+\tilde{\Omega}_{r,i}+\tilde{\Omega}_{\Lambda,i}\right),
\label{eq:OK_i}
\end{equation}
where the global (flat $\Lambda$CDM universe) $\Omega_{i}$ values
for each component is of course given by, e.g.
\begin{equation}
\Omega_{m,i}=\frac{\rho_{m,i}}{\rho_{{\rm crit},i}}=\frac{\Omega_{m,0}a_{i}^{-3}}{\Omega_{m,0}a_{i}^{-3}+\Omega_{r,0}a_{i}^{-4}+\Omega_{\Lambda,0}}.
\end{equation}

\subsection{Local cosmological parameters and redshift mapping\label{subsec:present_parameters}}

We take Enzo as our model simulation code, and show how cosmological
parameters are assigned when an overdense (underdense) patch is simulated.
Application to other codes will be similar. Because Enzo parameter
file requires those at the ``present'', such as \textsf{CosmologyOmegaMatterNow},
we need to get the local values of these. However, we can arbitrarily
define the ``present'', only if the local scale factor $\tilde{a}$
at that time, $\tilde{a}_{0}$, is normalized to 1. This convention
is used in Enzo. We note that the following assignment procedure is
provided as a separate function code in BCCOMICS.

Let us take some global redshift $z'$ and scale factor $a'=1/(1+z')$
to be those of the local present. We then need to connect this information
to the actual evolution of $\tilde{a}(t)$. We need to integrate $d\tilde{a}/dt$
(Equ. \ref{eq:h_hi})
\begin{eqnarray}
\frac{d\tilde{a}}{d(tH_{i})}=\tilde{a}_{i}\frac{\tilde{H}_{i}}{H_{i}}&&\left[\tilde{\Omega}_{m,i}\left(\frac{\tilde{a}}{\tilde{a}_{i}}\right)^{-1}+\tilde{\Omega}_{r,i}\left(\frac{\tilde{a}}{\tilde{a}_{i}}\right)^{-2}\right. \nonumber\\
&&\left. +\tilde{\Omega}_{\Lambda,i}\left(\frac{\tilde{a}}{\tilde{a}_{i}}\right)^{2}+\tilde{\Omega}_{K,i}\right]^{1/2},
\label{eq:dadt}
\end{eqnarray}
and then form a $\{ tH_{i},\,\tilde{a}(t)\}$ table. When performing
numerical integration in practice, we start from some time ($t_{s}$)
deep inside the radiation-dominated epoch, with $t_{s}H_{i}\ll1$,
and then use the analytical expression during this epoch 
\begin{equation}
\tilde{a}_{s}=\tilde{a}_{i}\sqrt{2t_{s}H_{i}\frac{\tilde{H}_{i}}{H_{i}}\sqrt{\tilde{\Omega}_{r,i}}},\label{eq:astart}
\end{equation}
where we set $\tilde{a}_{i}=a_{i}$ ``temporarily''. We find that
$t_{s}H_{i}=10^{-5}$ provides an excellent accuracy for post-recombination
epoch, and we use ODE45 (Runge-Kutta 4th order equivalent) of Matlab
and gnu octave, or simply the Simpson's rule with appropriate time-binning.
In either way we can easily obtain an accuracy of less than $10^{-4}$
at all $t$.

We then sample specific global time $tH_{i}$'s from the starting
(global) scale factor $a_{i}$ to the final (global) scale factor
$a'$, in terms of the global time variable $tH_{i}=\{t_{1}H_{i},\,t_{2}H_{i},\,...,\,t_{N}H_{i}\}$.
This becomes the set of $N$ epochs for data output. The corresponding
local scale factor of the patch will be $\tilde{a}=\{\tilde{a}_{1},\,\tilde{a}_{2},\,...,\,\tilde{a}_{N}\}$,
which can be obtained from the $\{ tH_{i},\,\tilde{a}(t)\}$  table. $\tilde{a}_{N}\ne a'$
in general. And for convenience, we match $\tilde{a}_{N}$ to the
local present. We thus rescale $\tilde{a}$ to $\tilde{\tilde{a}}=\left\{ \frac{\tilde{a}_{1}}{\tilde{a}_{N}},\,\frac{\tilde{a}_{2}}{\tilde{a}_{N}},\,...,\,1\right\} $,
and the corresponding redshifts become $\tilde{z}=\{\tilde{z}_{1},\,\tilde{z}_{2},\,...,\,0\}=\left\{ \left(\frac{\tilde{a}_{1}}{\tilde{a}_{N}}\right)^{-1}-1,\,\left(\frac{\tilde{a}_{2}}{\tilde{a}_{N}}\right)^{-1}-1,\,...,\,1-1\right\} $.
This list is assigned to \textsf{CosmologyOutputRedshift} in Enzo.
When necessary, one can always map $\tilde{z}$ to $z$ through the
$\{tH_{i},\, \tilde{a}(t),\, a(t)\}$ table.

We also need to change several other cosmological parameters. To obtain
\textsf{CosmologyOmegaMatterNow}, e.g., we calculate it by
\begin{eqnarray}
\tilde{\Omega}_{m,0}&=&\tilde{\Omega}_{m,i}\left(\frac{\tilde{a}_{N}}{\tilde{a}_{i}}\right)^{-3}\Bigg/
\left[\tilde{\Omega}_{m,i}\left(\frac{\tilde{a}_{N}}{\tilde{a}_{i}}\right)^{-3}+\tilde{\Omega}_{\Lambda,i}\right.\nonumber\\
&&\left. +\tilde{\Omega}_{r,i}\left(\frac{\tilde{a}_{N}}{\tilde{a}_{i}}\right)^{-4}+\tilde{\Omega}_{K,i}\left(\frac{\tilde{a}_{N}}{\tilde{a}_{i}}\right)^{-2}\right],
\label{eq:Om0}
\end{eqnarray}
and follow similar steps with correct power-law of
$(\tilde{a}_{N}/\tilde{a}_{i})$ in the numerator 
for all other \textsf{CosmologyOmega{*}Now}'s,
where {*}=\{\textsf{CDM}, \textsf{Baryon}, \textsf{Matter}, \textsf{Lambda},
\textsf{Radiation}\}. Non-zero curvature term for non-zero $\Delta_m$
should be taken in carefully, and in case of (recent-version) Enzo, this term is
calculated internally instead of being accepted as an input parameter.
\textsf{CosmologyHubbleConstantNow}, which is
the present Hubble constant in units of 100~${\rm km}\,{\rm s}^{-1}\,{\rm Mpc}^{-1}$, becomes
\begin{eqnarray}
\tilde{h}&=&\frac{\tilde{H}_{i}}{100\,{\rm km/s/Mpc}}\left[\tilde{\Omega}_{m,i}\left(\frac{\tilde{a}_{N}}{\tilde{a}_{i}}\right)^{-3}+\tilde{\Omega}_{\Lambda,i}\right.\nonumber\\
&&\left.+\tilde{\Omega}_{r,i}\left(\frac{\tilde{a}_{N}}{\tilde{a}_{i}}\right)^{-4}+\tilde{\Omega}_{K,i}\left(\frac{\tilde{a}_{N}}{\tilde{a}_{i}}\right)^{-2}\right]^{1/2}.
\label{eq:local_h}
\end{eqnarray}
In addition, because the comoving box size is the proper length at
``present'', this needs to be reassigned too. First, let us denote
the comoving length of the mean-density box by $L$. Then the proper
length of the box at $\tilde{a}_{i}$ is $L\tilde{a}_{i}$. The proper
length of the box at $\tilde{a}_{N}$, which is the comoving length
of the box, is then $L\tilde{a}_{i}\left(\tilde{a}_{N}/\tilde{a}_{i}\right)=L\tilde{a}_{N}$.
This is then assigned to\textsf{ CosmologyComovingBoxSize}. 

\subsection{Scaling laws and halo identification\label{subsec:scaling}}

Even when peaks and voids are treated as a separate universe, an observer
there can measure quantities based on the global properties of the
Universe. For example, an observer that estimated the local matter
content $\tilde{\Omega}_{m}$ through an observation inside a small
volume (e.g. only inside a 4 Mpc patch) will realize that the global
value $\Omega_{m}$ has been only underestimated after enlarging the
survey volume. Therefore, scaling laws for time and length are required.

The scaling laws are given trivially. If the local length and time
scales are $\tilde{L}$ and $\tilde{t}$, the global length scales
$L$ and $t$ are given by
\begin{equation}
L=\tilde{L}\frac{a}{\tilde{a}}\label{eq:scaling_length}
\end{equation}
and
\begin{equation}
T=\tilde{T}\frac{t}{\tilde{t}},\label{eq:scaling_time}
\end{equation}
respectively, where $t$ and $\tilde{t}$ are ages of the actual universe
and the local universe (patch), respectively.

A few obvious but important applications are imminent. If one were
to restrict the volume of a galaxy survey to e.g. $200^{3}\,{\rm Mpc}^{3}$
centered at the observer,
there would occur a danger of wrongfully measuring the BAO scale (true
value of $\sim$150 Mpc) as $150\,(\tilde{a}/a)$ Mpc where $\tilde{a}$
is the local scale factor of the $200^{3}\,{\rm Mpc}^{3}$ volume.
Similarly, the cosmological wavenumber of fluctuations should be scale
properly. If one conducted an auto-correlation analysis on a limited-volume
survey samples of galaxies, the local wavenumber $\tilde{k}$ and
the local correlation length $\tilde{l}$ should be mapped to the
global values as
\begin{equation}
k=\tilde{k}\frac{\tilde{a}}{a}\label{eq:wavenumber_scaling}
\end{equation}
and
\begin{equation}
l=\tilde{l}\frac{a}{\tilde{a}}.\label{eq:correlation_scaling}
\end{equation}
Equation (\ref{eq:wavenumber_scaling}) should be applied to the scaling
law suggested for the $k$-space fluctuation and the power spectrum
by \citet[Equs. 21 and 26]{Goldberg2004}, where they forgot to scale
the wavenumber.

Halo identification schemes should also be approached carefully, which
is of our keen interest. Let us restrict the discussion to one specific
halo identification scheme: the Friends-of-Friends (FoF) algorithm.
A halo is identified if a collection of particles are connected in
lengths that are smaller than the FoF linking length $b$, in units
of the mean particle separation. A given linking length $b$, is an
implicit indicator of the mean density of resulting halos $\left\langle \rho_{m}\right\rangle $:
($\left\langle \rho_{m}\right\rangle \simeq180(b/0.2)^{-3}\bar{\rho}_{m}$
or $\delta_{{\rm lin}}>\delta_{{\rm crit}}\sim1.67$, as in \citealt{Lacey1994}).
Because the relation
\begin{equation}
\frac{\left\langle \rho_{m}\right\rangle }{\bar{\rho}_{m}}\simeq180\left(\frac{b}{0.2}\right)^{-3}\label{eq:fof}
\end{equation}
roughly holds regardless of cosmology for any simulation box (\citealt{Lacey1994}),
this can be interpreted in our case as
\begin{equation}
\frac{\left\langle \tilde{\rho}{}_{m}\right\rangle }{\bar{\tilde{\rho}}_{m}}\simeq180\left(\frac{\tilde{b}}{0.2}\right)^{-3}\label{eq:fof_local}
\end{equation}
for any local patch, where halos with a \emph{local} linking length
$\tilde{b}$ in units of the \emph{local} mean particle separation
will have the \emph{local} overdensity $180\left(\tilde{b}/0.2\right)^{-3}$.
Now, any halo inside our universe is defined in terms of the global
mean density $\bar{\rho}_{m}$ to make $\left\langle \tilde{\rho}{}_{m}\right\rangle /\bar{\rho}_{m}=180$$\left(b/0.2\right)^{-3}$,
and thus
\begin{equation}
\tilde{b}=b\left(\frac{a}{\tilde{a}}\right).\label{eq:b_local}
\end{equation}
Equation (\ref{eq:b_local}) is indeed consistent with a simple fact
that a single proper linking length, $b\times$(global mean particle
separation), should be applied universally if one imagines a very
large, mean-density simulation box that encloses many overdense and
underdense patches. 

Once a local patch is treated as a separate universe, then FoF halos
there should be generated using $\tilde{b}$ given by equation (\ref{eq:b_local}),
which requires scaling $\tilde{b}$ in a time-dependent way if a constant
$b$, e.g. $b=0.2$, is used. The minimum number of N-body CDM particles
for halo identification need not change, because a universal criterion
should be used. For example, let us imagine a certain overdense patch
with $\tilde{a}=0.05$ but $a=0.1$ at the same cosmic time $tH_{i}=10$.
If the ``usual'' linking length is 0.2 of the mean particle separation
in the mean-density patch, then the linking length of the overdense
patch should be $0.2\times(0.1/0.05)=0.4$ of the ``local'' mean
particle separation.

Finally, if some sort of overdensity threshold is used, e.g. $\delta_{{\rm th}}=1$
for triggering a certain astrophysical process, this needs to be scaled
too. To have a universal density threshold, $\rho_{{\rm th}}=(1+\tilde{\delta}_{{\rm th}})\tilde{\rho}_{b}=(1+\delta_{{\rm th}})\rho_{b}$,
we need $\tilde{\delta}_{{\rm th}}=(\tilde{a}/a)^{3}(1+\delta_{{\rm th}})-1$.
If a threshold is given instead in terms of $n_{{\rm th}}$ or $\rho_{{\rm th}}$
because a physical process of interest is dependent on the proper
density, then there is no need for scaling.

\section{Application\label{sec:Application}}

As an application, we use BCCOMICS for generating initial conditions
and perform a suite of cosmological simulations of structure formation
using Enzo (\citealt{Bryan2014}), sampling a few patches of varying
$\Delta_{{\rm c}}$ and $V_{{\rm bc}}$. One can consider other large-scale
variables ($\Theta_{{\rm c}}$, $\Delta_{{\rm b}}$, $\Theta_{{\rm b}}$,
and $\Delta_{T}$) as well. However, strong correlation between $\Theta_{{\rm c}}$
and $\Delta_{{\rm c}}$ alleviates the need for considering $\Theta_{{\rm c}}$,
and \{$\Delta_{{\rm b}}$, $\Theta_{{\rm b}}$, $\Delta_{T}$\} are
of less importance than $\Delta_{{\rm c}}$ and $V_{{\rm bc}}$. The
rather quick transition of a very loose correlation between $\Delta_{{\rm c}}$
and $\Delta_{{\rm b}}$ at $z=1000$ to a very strong correlation
at $z=200$ is one of the reasons for ignoring $\Delta_{{\rm b}}$
in this work. Nevertheless, one should be wary of the separate impact
of $\Delta_{{\rm b}}$ on the ever-existing BAO feature in the matter
density fluctuation $\Delta_{+}$ and a subsequent impact on the small-scale
structure formation, if e.g. some type of cosmology with first galaxies
is considered (e.g. \citealt{McQuinn2012}). The set of large-scale
variables are \{$\Delta_{{\rm c}}$, $V_{{\rm bc}}(z=1000)/({\rm
  km}\,{\rm s}^{-1})$\}=\{0,
0\}, \{0, 26.5\}, \{$2\sigma_{\Delta_{{\rm c}}}$, 26.5\}, and \{$2\sigma_{\Delta_{{\rm c}}}$,
38\}, which are sampled over 4-Mpc patches at $z=1000$. $V_{{\rm bc}}=0$ case
becomes too rare to be realized in our $(604\,{\rm Mpc})^{3}$ box
when $\Delta_{{\rm c}}=2\sigma_{\Delta_{{\rm c}}}$ (see Fig \ref{fig:z1000}f).
These sampling parameters are listed in Table \ref{tab:cases} with
case names. 

\begin{table}
\begin{tabular}{|c|c|c|c|c|}
\hline 
 & D0V0 & D0V2 & D2V2 & D2V3\tabularnewline
\hline 
\hline 
$\Delta_{{\rm c}}$ & 0 & 0 & $2\sigma_{\Delta_{{\rm c}}}$ & $2\sigma_{\Delta_{{\rm c}}}$\tabularnewline
\hline 
$V_{{\rm bc}}$~(km/s) & 0 & 26.5 & 26.5 & 38\tabularnewline
\hline 
\end{tabular}

\caption{All cases use initial conditions with a universal random seed, a grid
resolution and the CDM-particle number of $512^{3}$ each, and the
initial box size of 1 Mpc. High density cases D2V2 and D2V3 gradually
detaches from the Hubble expansion and thus the comoving size of these
boxes shrinks in time, in contrast to the cases D0V0 and D0V2.\label{tab:cases}}

\end{table}

For fair comparison, in generating small-scale fluctuations, we apply
a single universal Gaussian random seed to all these cases. Grid quantities
are generated on a uniform grid of $512^{3}$ cells, and dark matter
particle displacements are based on uniform spacing. Even though the
large-scale quantities are sampled over 4-Mpc patches, we let simulation
boxes to be of 1 Mpc to resolve minihalos down to $M\sim 10^{5}\,M_{\odot}$.
Chemistry of and cooling by primordial elements (H, He and their ions)
are calculated, and formation of Pop III stars are tracked by sink
particles when the number density criterion for baryons, $n_{b}>10^{3}\,{\rm cm^{-3}}$,
is met. This is still a pilot study, and we do not calculate the radiation
transfer and its effects on gas, and turn off the mesh refinement. The subsequent
paper (Paper II, in preparation) will adopt a more aggressive configuration
that is suitable for the study of first star formation.

We stress this fact again: it is not a good practice to assume $V_{{\rm bc}}=0$
or impose a sudden $V_{{\rm bc}}$ (as is done by many simulations:
\citealt{Greif2011,Maio2011,Stacy2011,Schauer2017,Hirano2017} ) in
small-scale structure formation simulations, which is one of a few
reasons why one should use at least CICsASS for $\Delta_{{\rm c}}=\Delta_{{\rm b}}=0$
case or BCCOMICS for more generic cases of non-zero $\Delta_{{\rm c}}$
and other large-scale quantities. First, $V_{{\rm bc}}=0$ lies at
the very end of the Maxwell-Boltzmann distribution tail (Equ. 14 of
Ahn16) and thus is too a rare event. Second, a sudden imposition of
non-zero $V_{{\rm bc}}$ on an initial condition based on CAMB transfer
functions will underestimate the large-scale environmental effect ($V_{{\rm bc}}$
only for CICsASS, and all possible variants for BCCOMICS) on small-scales,
which has continued since the recombination epoch (see also the discussion
by \citealt{O'Leary2012}). CAMB and other Boltzmann solvers do not
solve equation (\ref{eq:TH}), not to mention equation (\ref{eq:Ahn}),
and thus a sudden imposition of $V_{{\rm bc}}$ onto an initial condition
based on these Boltzmann solvers suffers from this problem. 

\begin{figure*}
\includegraphics[width=\textwidth]{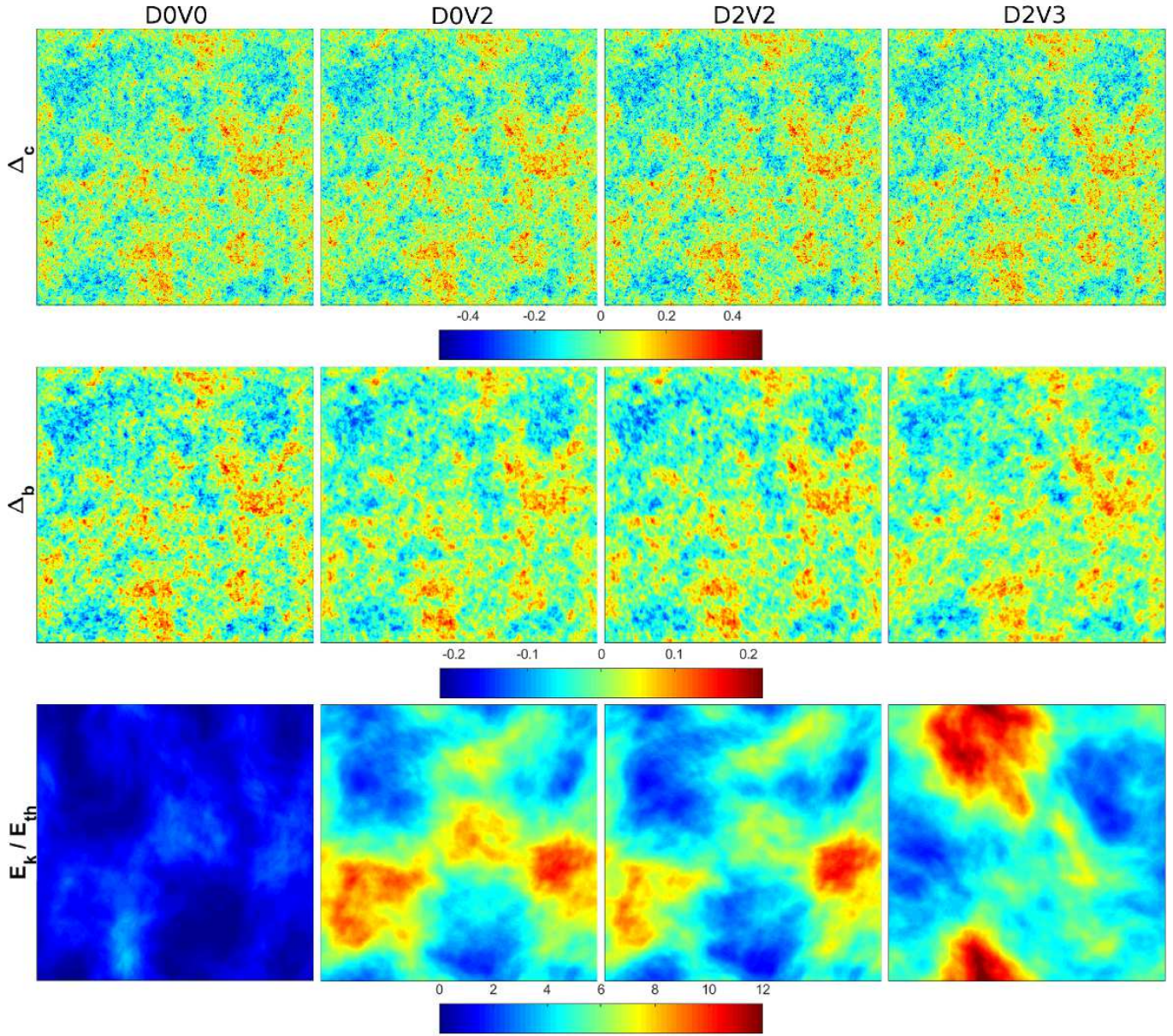}

{} 

\caption{Overdensity maps of CDM (top) and baryons (middle), and the kinetic-to-thermal
energy ratio of baryons (bottom) at $z=200$, generated by BCCOMICS.
From left to right, the CDM overdensity and streaming-velocity environments
are \{$\Delta_{{\rm c}}$, $V_{{\rm bc}}$\}=\{0, 0\} (D0V0), \{0,
26.5~km/s\} (D0V2), \{$2\sigma_{\Delta_{{\rm c}}}$, 26.5~km/s\}
(D2V2), and \{$2\sigma_{\Delta_{{\rm c}}}$, 38~km/s\} (D2V3), where
$V_{{\rm bc}}=V_{{\rm bc}}(z=1000)$. Note that directions of ${\bf V}_{\rm bc}$ are 
not identical, and neither are directions of the baryon-density smear. \label{fig:IC_z200}}
\end{figure*}

We show three types of field maps at z=200 in Fig. \ref{fig:IC_z200}, which
are CDM overdensity, baryon overdensity and the ratio of kinetic-to-thermal
energies of baryons, under varying overdensity and streaming-velocity 
environments. In this figure, CDM overdensity maps look indistinguishable
from one another, because the dynamics of CDM is dominated by the
self-gravity of CDM, while baryon overdensity maps show distinguishable
smear that gets stronger as $V_{{\rm bc}}$ increases. The ratio of
kinetic-to-thermal energies of baryons overall increases as $V_{{\rm bc}}$
increases. If we imagine the practice of a sudden imposition of $V_{{\rm bc}}$,
the baryon overdensity maps would not show any mutual difference.
The relative importance of $V_{{\rm bc}}$, in terms of energetics,
also gets stronger as $V_{{\rm bc}}$ increases, as seen in figures
of the energy ratio. 

\begin{figure*}
\includegraphics[width=0.45\textwidth]{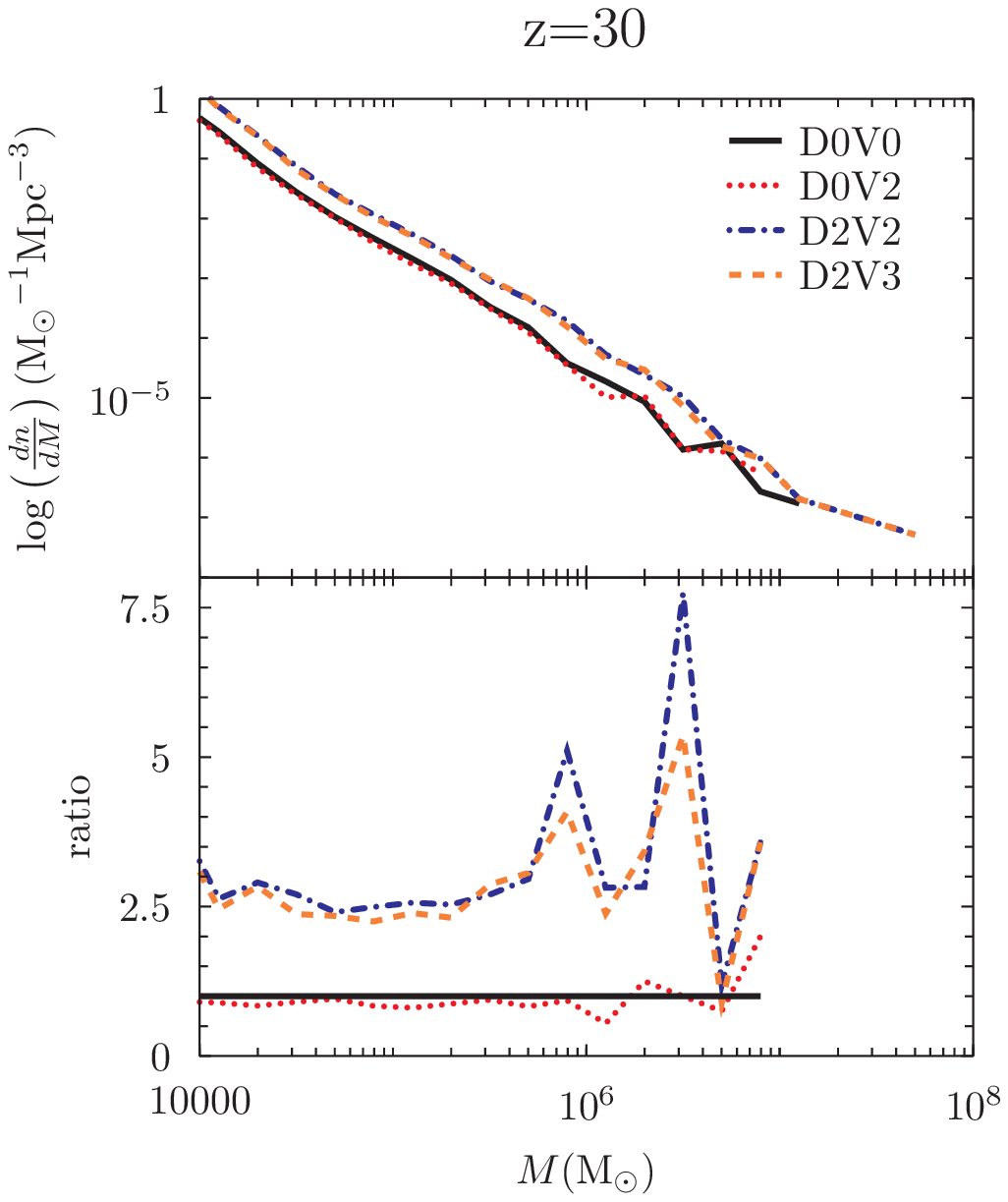}\includegraphics[width=0.45\textwidth]{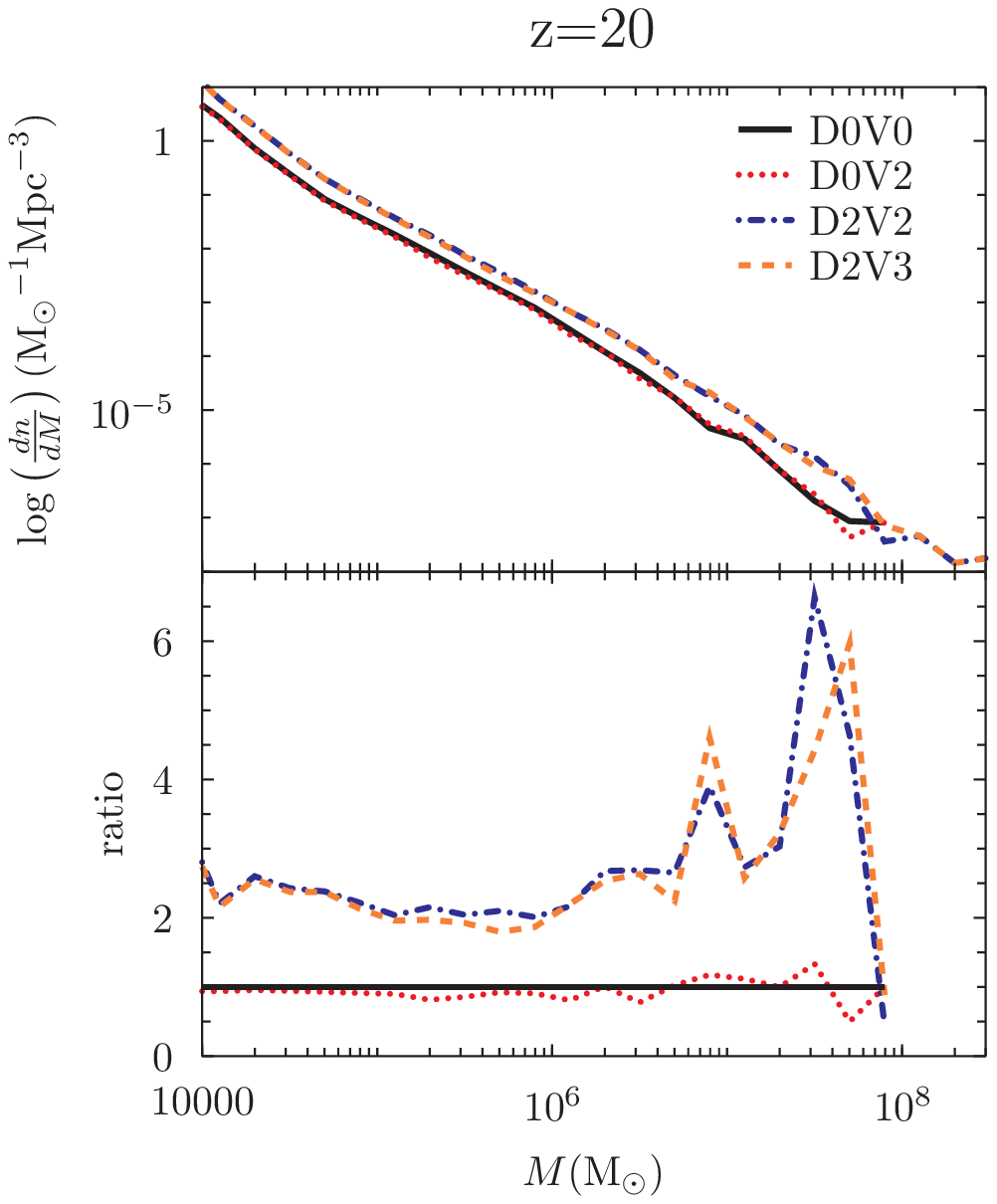}

\caption{Halo mass functions under different \{$\Delta_{{\rm c}}$, $V_{{\rm bc}}$\}
environments (D0V0: black solid; D0V2: red dotted; D2V2: blue dotted-dashed;
D2V3: orange dashed), obtained from N-body+hydro simulations
based on initial conditions at $z=200$ depicted in Fig. \ref{fig:IC_z200}.
Plotted are halo mass functions (upper) and the ratios with respect
to D0V0 case (lower) at $z=30$ (left) and $z=20$ (right).\label{fig:hmf}}
\end{figure*}

After simulating structure formation based on initial
conditions generated by BCCOMICS, we identified halos using the FoF scheme. As described
in Section \ref{subsec:scaling}, a universal linking length $b=0.2$
was used for all cases. For overdense patches (D2V2, D2V3), this translates
to the local linking length $\tilde{b}=0.2(a/\tilde{a})$. We used
\texttt{yt}\footnote{http://yt-project.org} analysis tool for halo
identification, and because this tool is keen only to local values
including the local mean particle separation, we fed this time-varying
$\tilde{b}(z)$ into\texttt{ yt} for any redshift $z$. Figure \ref{fig:hmf}
is the halo mass function $dn/dM$ for the net DM mass $M$ of FoF
halos. 

\begin{figure*}
\gridline{
\fig{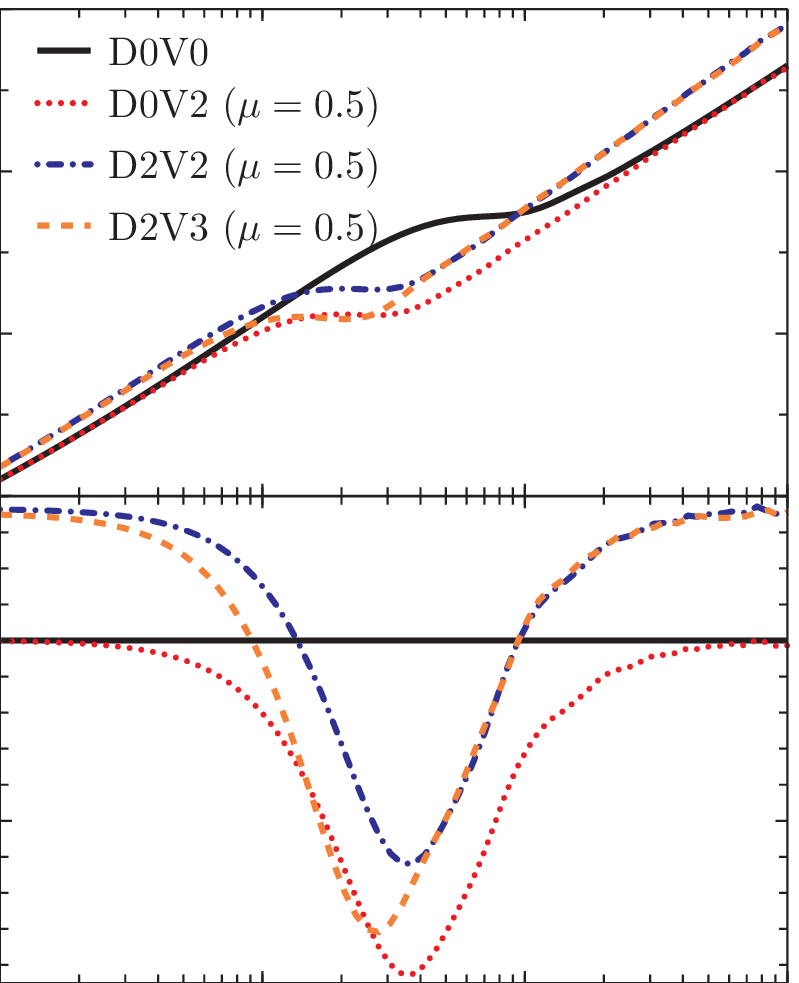}{0.4\textwidth}{(a)}
\fig{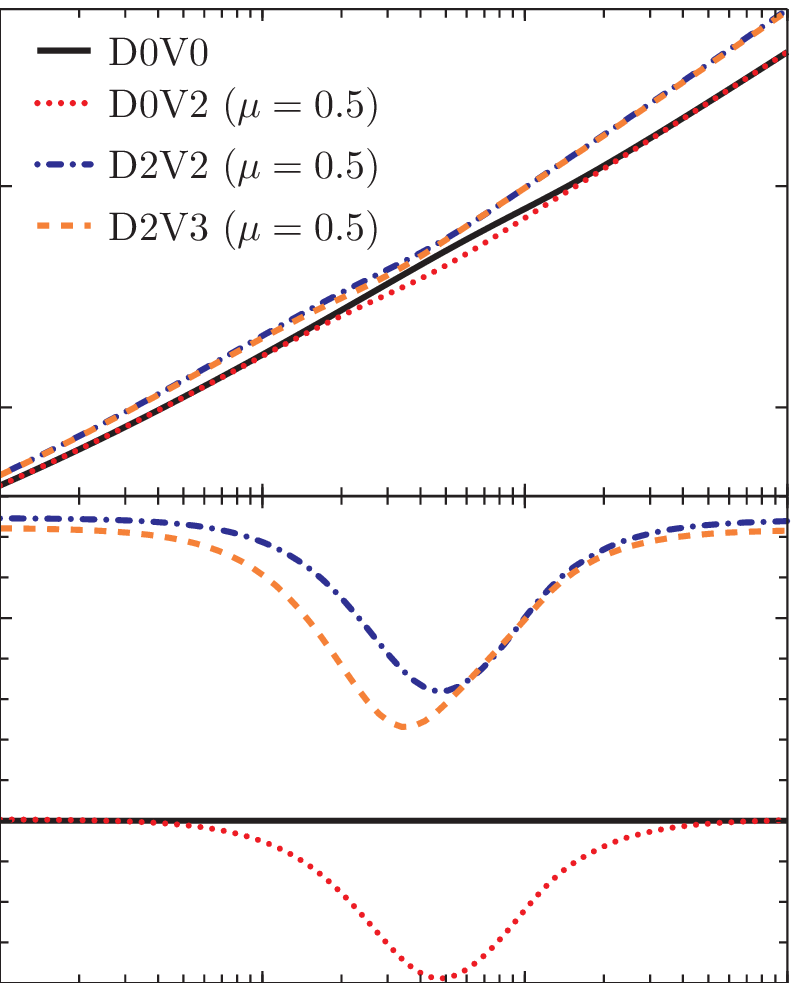}{0.4\textwidth}{(b)}
}
\caption{(a) Matter power spectrum at $z=200$, expressed in
terms of the ${\bf k}$-space variance $\Delta_{m}^{2}$, for different
patches (D0V0: black solid; D0V2: red dotted; D2V2: blue dotted-dashed;
D2V3: orage dashed). Except for the D0V0 patch, a constant 
$\mu\equiv \cos({\bf k},{\bf V}_{\rm bc})=0.5$ cases are selected
for other patches for a fair comparison of the negative impact of
$V_{\rm bc}$. 
The ratio in the bottom is with respect to the D0V0 case.
(b) Same as (a), but for the CDM power spectrum.
\label{fig:kD2}}
\end{figure*}

A few features are notable. First, overdense patches contain
more halos across the full mass range than mean-density patches, as
expected. Second, when overdensity environment is the same, higher
$V_{{\rm bc}}$ yields a stronger suppression. Third, the impact of
$V_{{\rm bc}}$ weakens in time (see bottom panels of Fig. \ref{fig:hmf}),
which is expected from the fact that $V_{{\rm bc}}\propto1/a$. These
features are consistent with the positive effect of overdensity and
the negative effect of $V_{{\rm bc}}$ on the clumping of baryons
inside DM halos  and even DM clumping itself, as seen in the ${\bf
 k}$-space variance of 
matter (CDM+baryon), $\Delta_{m}^{2}\equiv P_{m}(k)k^{3}/(2\pi^{2})$,
and that of CDM, $\Delta_{\rm c}^{2}\equiv P_{\rm c}(k)k^{3}/(2\pi^{2})$.
Fig. \ref{fig:kD2} shows that $\Delta_{\rm c}$ boosts both $P_{m}(k)$ 
and $P_{\rm c}(k)$ at
all values of $k$, while $V_{\rm bc}$ suppresses both $P_{m}(k)$ 
and $P_{\rm c}(k)$ in a
bound region of $k$.
Aside from very rare and massive halos ($M\gtrsim10^{7}\,M_{\odot}$)
we could identify until $z\simeq20$, the number density of halos
are $\sim${[}2-4{]} times as high as that in mean-density patches.
Because we allowed gas cooling, it is probable that the negative effect
of $V_{{\rm bc}}$ is somewhat reduced compared to the case without
cooling (e.g. \citealt{O'Leary2012}). One subtle feature is that
suppression of halo formation is not biased toward the low-mass
end. TH predicts that at $z\sim 40$, halos with mass $M\sim 10^{6}
M_{\odot}$ will be  more strongly suppressed than those with e.g. 
$M\sim 10^{4} M_{\odot}$. In Fig. \ref{fig:hmf}, this tendency can be
barely observed. For high-mass end, however, such a comparison is not
meaningful due to poor statistics. We plan to obtain better
statistics in Paper II by enlarging the simulation box size (including
larger-scale modes of fluctuation, equivalently) and thus
allowing more frequent formation of massive halos.

\begin{figure}
\includegraphics[width=0.4\textwidth]{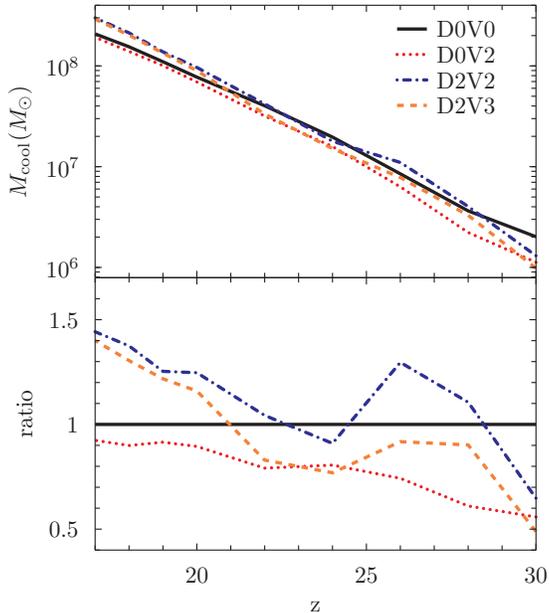}

\caption{Evolution of the total cooling mass inside the simulation box (top) and the
ratio to the D0V0 case (bottom). The line-type convention
is the same as Figure \ref{fig:hmf}. Note that this figure should be considered only for
a qualitative comparison, because the numerical resolution of these
simulations,  $512^{3}$ uniform-grid on a simulation box of volume
  $(1\,h^{-1}\,{\rm Mpc})^3$,
is currently not adequate for {\bf accurately} predicting this quantity.\label{fig:cool_gas} }
\end{figure}

The amount of gas that is first gravitationally bound in halos or filaments and
then undergoes cooling is closely related to the star formation process.
Because we did not allow adaptive mesh refinement (AMR), we defer our analysis on the star formation
to Paper II but instead analyzed the total amount of cooling gas.
We define the cooling gas by the criterion $t_{{\rm cool}}<t_{{\rm dyn}}$,
where $t_{{\rm cool}}$ and $t_{{\rm dyn}}$ are the local cooling
time and the local dynamical time, respectively. Figure \ref{fig:cool_gas}
shows the evolution of the total amount of cooling gas ($M_{\rm cool}$) in each simulation box.
We can take this quantity as 
a rough indicator of star formation activity. As nonlinear structures
grow in time, $M_{\rm cool}$ increases in all cases.
The positive effect of overdensity and the negative effect of $V_{{\rm bc}}$
on gas clumping (or cooling) is also clearly observed. 
In the early phase, at $z \sim 30$, D0V0 case has the largest $M_{\rm cool}$ 
because the density environments of D2V2 and D2V3 have not deviated too much
from those of D0V0 and D0V2, and therefore the hierarchy roughly
follows that {\bf of} $V_{{\rm bc}}$
in descending order: $M_{\rm cool}({\rm D0V0})> M_{\rm cool}({\rm D0V2})\simeq M_{\rm cool}({\rm D2V2})>M_{\rm cool}({\rm D2V3})$.
After that, however,
the positive effect of overdensity takes over as density peaks are more
detached from the global Hubble flow, and D2V2 case becomes the site for
the most efficient cooling. At $z\lesssim 20$, the positive effect from density dominates 
over the early-phase negative effect from $V_{{\rm bc}}$, but still retains
the memory of the negative effect from $V_{{\rm bc}}$, such that $M_{\rm cool}({\rm D2V2})> M_{\rm cool}({\rm D2V3}) > M_{\rm cool}({\rm D0V0})>M_{\rm cool}({\rm D0V2})$.
This is indeed an epoch where minihalos provide a significant contribution to cosmic
reionization (e.g. \citealt{Ahn2012}), and thus our study is expected to improve
upon the existing scenarios of reionization.

We will extend this application with a more self-consistent treatment in Paper II, 
especially with the AMR capability on. The quantitative result of this section, therefore,
is likely to be changed. Nevertheless, the qualitative result is expected to remain the same.

\section{Summary\label{sec:Summary-and-Discussion}}

First stars and first galaxies are created inside minihalos, which
are the first nonlinear structure in the history of the universe.
The evolution of small-scale fluctuations of both CDM and baryons
are found strongly affected by the large-scale environment, leading
to cosmic variance in the formation and evolution of these first objects.
This effect is caused predominantly by the large-scale density and
the baryon-CDM streaming velocity. Because this effect has not been
fully incorporated in existing initial condition generators, we have
developed an initial condition generator, BCCOMICS, that fully incorporates
this effect for the first time. BCCOMICS first calculates the evolution
equations, which were given by TH for only the mean-density environment
and by Ahn16 for a fully generic, non-zero overdensity environment,
and then generates three-dimensional fields of grid and particle quantities.

Study of this cosmic variance requires realizing the local environment
in simulations. This can be realized as zoom-in patches in one big
simulation box, or as an individual patch with a periodic box condition.
For the latter, we have developed a systematic scheme to simulate
the growth of small-scale structure, inside density peaks and voids,
by treating the environment as a separate universe with local cosmological
quantities. The affected quantities are the local Hubble parameter,
the local cosmic abundance of various contents (CDM, baryon, radiation,
cosmological constant), and the local scale factor. Analysis of simulation
data requires a scaling law that maps the local quantity to the global
quantity. For example, a correlation length of galaxy clustering or
the spatial BAO peak of the correlation function, which are found
through a local galaxy survey inside an overdense environment, will
differ from the global values found through a unlocalized galaxy survey.
We provided a trivial but important scaling law that allows one to
easily deduce the corresponding global quantities.

As a pilot study, we generated initial conditions by BCCOMICS and
performed a suite of N-body+hydro simulations of small-scale
structure formation under varying large-scale environments. As expected,
the overdensity environment yields positive feedback effects and the
streaming-velocity environment negative feedback effects. Compared
to the mean-density environment, halos are generated in higher population
and gas cooling becomes more efficient in the overdense environment.
The higher the overdensity is, the stronger this positive feedback will
become. In contrast, the streaming velocity tends to suppress halo
formation and also the gas cooling. 

We need to improve upon 
the quantitative prediction of this pilot study, and will conduct a more self-consistent study
by allowing AMR and possibly the radiation transfer as well. At the same time,
a wider parameter space of environmental variation will be considered, and
its result will be described in Paper II (in preparation).

\begin{widetext}
\acknowledgments
We thank the anonymous referee for a clear report. This work was supported by the NRF grant 2016R1D1A1B04935414 and a research grant from Chosun University (2016). 
\end{widetext}

\bibliographystyle{aasjournal}

\end{document}